\documentclass[fleqn,usenatbib]{mnras}

\usepackage{newtxtext,newtxmath}
\usepackage{makecell}
\usepackage[T1]{fontenc}

\DeclareRobustCommand{\VAN}[3]{#2}
\let\VANthebibliography\thebibliography
\def\thebibliography{\DeclareRobustCommand{\VAN}[3]{##3}\VANthebibliography}

\usepackage{graphicx}	
\usepackage{amsmath}	
\usepackage{natbib}
\usepackage{xcolor}

\title[2021 Outburst of GX339-4 with \textit{AstroSat}]{Evolution of the 2021 Outburst of GX 339-4 with \textit{AstroSat}}

\author[V. Sharma et al.]{
Vaibhav Sharma,$^{1}$\thanks{E-mail: svbhv@iitk.ac.in and vaibhavsharmaiitk@gmail.com}
Ranjeev Misra,$^{2}$
J S Yadav$^{3,4}$
Akash Garg$^{2}$
Pankaj Jain$^{1,3}$
\\
$^{1}$Department of Physics, Indian Institute of Technology Kanpur, Kanpur Nagar, Uttar Pradesh 208016, India\\
$^{2}$Inter-University Center for Astronomy and Astrophysics, Ganeshkhind, Pune, Maharashtra 411007, India\\
$^{3}$Space, Planetary and Astronomical Sciences and Engineering, IIT Kanpur, Kanpur Nagar, Uttar Pradesh 208016, India\\
$^{4}$Department of Astronomy and Astrophysics, Tata Institute of Fundamental Research, Mumbai, Maharashtra 400005, India
}

\date{Accepted 2025 December 22. Received 2025 December 19; in original form 2025 June 21}

\pubyear{\the\year{}}

\begin{document}
\label{firstpage}
\pagerange{\pageref{firstpage}--\pageref{lastpage}}
\maketitle

\begin{abstract}
We present a comprehensive study of the 2021 outburst of GX 339-4 using \texttt{AstroSat} observations in the hard-intermediate (HIMS) and soft-intermediate states (SIMS). Spectral and timing analyses across these states suggest that during the SIMS, unabsorbed flux (0.1-3 keV), inner disc temperature, and "apparent" inner disc radius do not change, suggesting the stability of the disc. In the SIMS, the photon index decreases from $\sim 2.1$ to $\sim 1.7$, indicating spectral hardening. The power density spectra (PDS) suggest the presence of quasi-periodic oscillations (QPOs) in the HIMS and SIMS. The QPO frequency evolves from $\sim0.1$ Hz to $\sim0.2$ Hz in the HIMS, and further to $\sim5.7$ Hz in the SIMS. We also observe a decrease in QPO frequency from $\sim5.7$ Hz to $\sim 4.5$ Hz during the SIMS. We discuss the evolution of the QPO, fractional root mean square (rms) amplitude, and time-lag spectra. We discover that variations in disc normalization, disc temperature, and coronal heating rate can reproduce the observed rms and lag spectra with a time delay between them.

\end{abstract}




\section{Introduction}
X-ray binaries (XRBs) are systems consisting of a compact object and a companion star (or donor) orbiting around the center of mass of the combined system. Based on the nature of the compact object, the XRBs are classified into two categories. If the compact object is a neutron star, the system is referred to as a neutron star X-ray binary (NSXB); if the compact object is a black hole, the system is called a black hole X-ray binary (BHXB). The XRBs are further classified into two categories based on the mass of the companion star. If mass of the donor star is $\leq \textup{M}_\odot$, the system is known as a low-mass X-ray binary (LMXB), and if mass of the donor star is $\geq 10 \, \textup{M}_\odot$, the system is classified as a high-mass X-ray binary (HMXB).

In general, the black hole (BH) LMXBs exhibit frequent outbursts due to their transient nature. They spend most of their time in quiescence, followed by an outburst that last from a few days to several months \citep[e.g.][]{Dubus2001,Deegan2009}. A primary cause of the outburst is the change in mass accretion rate, which transitions the source from the quiescence phase to the outburst phase \citep[e.g.][]{Done2007}. During an outburst, both spectral states and temporal properties change. Most of the BH LMXBs follow a q-shaped track in the intensity versus hardness ratio plot during an outburst, commonly known as hardness-intensity diagram (HID; \cite{Homan2001}; \citealt{Belloni2005, HomanBelloni2005}), while \cite{Tetarenko2016} and \cite{Alabarta2021} reported that 30-40 $\%$ of the total BH LMXBs do not follow the q-shaped track in the HID during their outburst. The intensity in the HID is typically from both thermal and non-thermal emissions, while the hardness ratio (HR2) is defined as the ratio between the intensity of the source in two energy bands. We define it as the ratio between the intensity of the source in 15-50 keV and 2-20 keV. At the beginning of the outburst, the BH LMXBs are observed in low/hard state (LHS), during which the non-thermal emission dominates. As the outburst progresses, the BH LMXBs transition from the LHS to high/soft state (HSS) through intermediate state (IMS) \citep{Remillard2006}. In the HSS, the thermal emission from the accretion disc dominates. The IMS are further categorized into two states: the soft intermediate state (SIMS) and hard intermediate state (HIMS).

X-ray energy spectrum of a BH LMXB can be described by a multi-color disc blackbody and a power law. The multi-color disc blackbody emission is believed to originate from a geometrically thin and optically thick accretion disc \citep{Shakura1973}. A fraction of disc photons is inverse Comptonized by highly energetic electrons, called the corona, near the BH, producing high-energy tail in the energy spectrum described by the power law \citep{Haardt1993, Chakrabarti1995, Done2007}. The inverse Comptonized photons may be reprocessed in the disc, producing iron K$_{\alpha}$ line around $6.4$ keV. Additionally, we may observe a reflection hump at $\sim 20-40$ keV, resulting from the reprocessing of inverse Comptonized photons in the disc \citep{Fabian1989, Matt1991}.

The BHXBs also exhibit X-ray variability in the light curve, which can be seen in the power density spectrum (PDS) as a broad-band noise and sometimes a narrow peaked feature, popularly known as quasi-periodic oscillation (QPO) \citep{Nowak2000, Belloni2002, Remillard2006, Ingram2019}. QPOs are classified as low-frequency QPOs (LFQPO) with frequencies less than $30$ Hz and high-frequency QPOs (HFQPO) with frequencies ranging from $30$ Hz to a few hundred Hz \citep{Belloni2012, Mendez2013, Motta2022}. The LFQPOs are commonly classified into three types—A, B, and C \citep{Wijnands1999, Sobczak2000, Casella2005, Motta2015, Ingram2019}. Based on their QPO frequency ($\nu_{0}$), fraction root mean square (rms) amplitude, and quality factor (Q), these types can be characterized as follows: type A ($\nu_{0}$ $\sim8$ Hz, $\leq 3\%$ rms, Q $\leq 3$), type B ($\nu_{0}$ $\sim5-6$ Hz, $\sim 2-4\%$ rms, Q $\geq 6$), and type C ($\nu_{0}$ $\sim0.1-15$ Hz, $\sim 3-16 \%$ rms, Q $\sim 7-12$) \citep{Casella2005,Ingram2019}. The rms amplitude of the QPO is defined as the square root of the normalization of fitted \texttt{Lorentzian} component to the QPO, assuming that the PDS is rms-normalized. The quality factor (Q) is defined as the ratio of the QPO frequency $\nu_{0}$ to its full width at half maximum (FWHM).
\begin{figure}
	\includegraphics[width=\columnwidth, height=5.2cm]{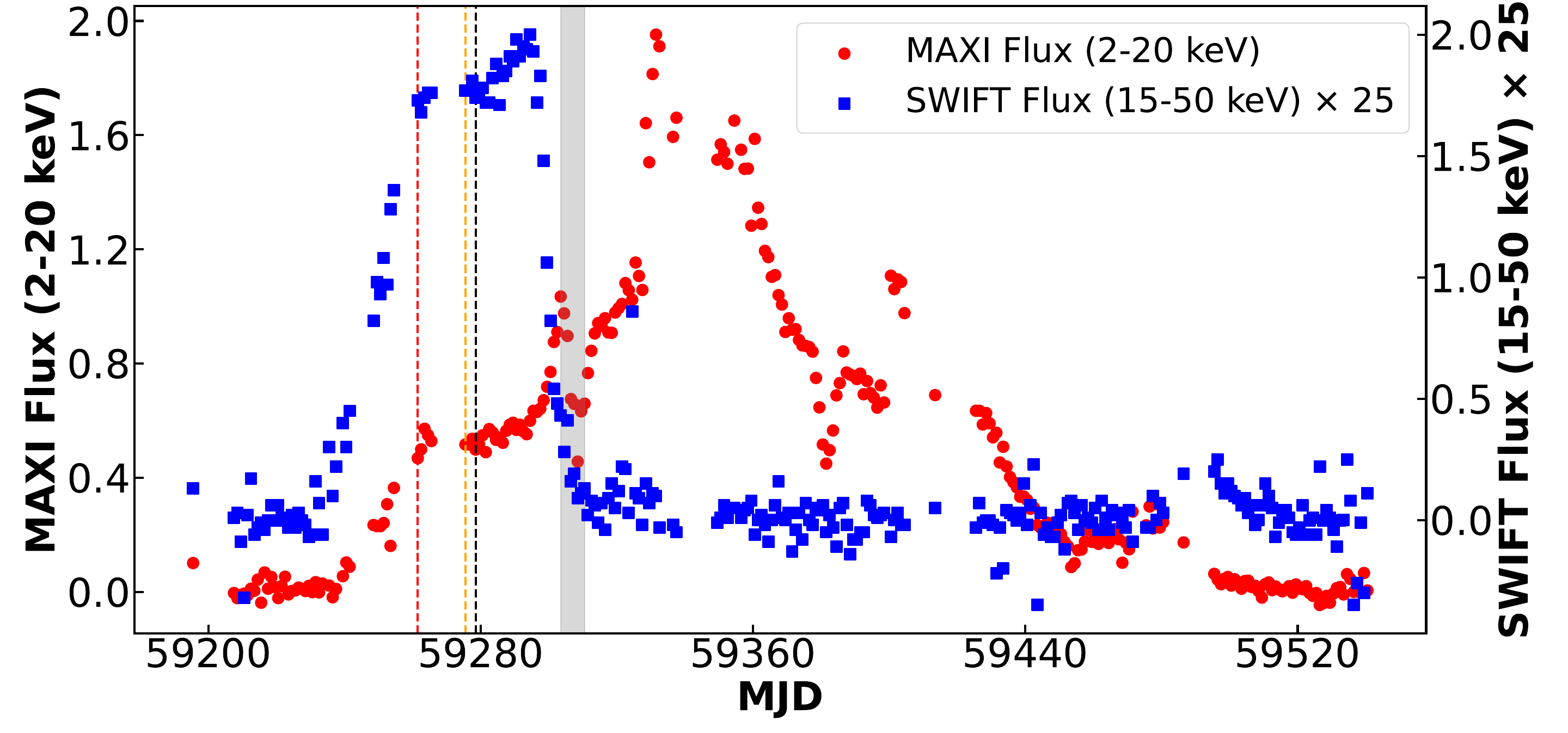}
    \caption{MAXI (2.0–20.0 keV) and Swift/BAT (15.0–50.0 keV) flux light curves of GX 339-4 during its 2021 outburst. The red, orange, and black vertical dashed lines, as well as the dashed grey region indicate the AstroSat observation period. The Swift/BAT flux is scaled by a factor of 25 for comparison.}

    \label{fig: MAXI_SWIFT_LC}
\end{figure}

Extensive studies have been conducted to investigate the origin of QPOs, but it remains a topic of intense research due to the lack of consensus. Many models have been proposed to explain the origin of QPOs, among which several models consider a geometric origin. \citet{Stella1997} and \citet{Stella1999} discuss the relativistic precession model (RPM), which suggests the Lense-Thirring precession as the model for QPO. \citet{Ingram2009} considered the precession of the hot inner flow around the truncated disc \citep{Done2007} and described this precession as the QPO frequency. A few models, such as \citet{Titarchuk2000}, \citet{Chakrabarti2008}, \citet{Karpouzas2020}, and \citet{Bellavita2022}, assume that the QPOs are intrinsic to the accretion flow, arising from instabilities within the accretion flow that lead to quasi-periodic variations in the observed X-ray flux.

The QPO properties evolve as the outburst of the BH LMXB progresses (e.g. \citealt{Ingram2019}), and they are tightly correlated with the different spectral states. This suggests that as the spectral properties of the source evolve, the QPO properties also change, indicating correlation between the spectral and temporal properties. \citet{Misra2013} discuss a general model to explain the origin by introducing small variations in the spectral properties that characterize the energy spectrum. \citet{Maqbool2019} fit the rms and time-lag spectra of Cyg X-1 with a stochastic propagation model, defined by three parameters: inner disc radius, disc temperature at the truncated radius, and time lag between them.

\citet{Garg2020} also develop a generic model to investigate the responsible radiative component for energy-dependent properties of the QPOs by introducing small amplitude variations in the physical spectral parameters. \citet{Garg2020} assume that the energy spectrum is composed of two components: black body radiation from the accretion disc and inverse Comptonization of the disc photons from the corona. \citet{Garg2022} apply the same model to the QPOs found in MAXI J1535-571, fitting the rms and time-lag spectra by introducing variations in the mass accretion rate ($\dot m$), inner disc radius (R$_{in}$), and coronal heating rate ($\dot H$). Using this model, \citet{Nazma2023, Arbind2025} reproduce the rms and time lag spectra of the QPO observed in H 1743-322 and MAXI J1803-298, while \citet{Dhaka2024} and \citet{Hitesh2024} fitted the energy-dependent rms and time-lag of the broad feature observed in GRS 1915+105 and GX 339-4, respectively.

BH LMXB GX 339-4 is a well-known source for its frequent outbursts. It is discovered in 1973 by the MIT X-ray detector onboard OSO-7 \citep{Markert1973}, and since then, it has been found to undergo frequent outbursts approximately every 2-3 years. The mass of BH (M$_{BH}$) in GX 339-4 is still debated, with recent estimates ranging from $7.8-10.6 M_\odot$ \citep{Parker2016}, $2.3-9.5 M_\odot$ \citep{Heida2017}, to $8.28-11.89 M_\odot$ \citep{Sreehari2019}. \citet{Hynes2003} and \citet{Heida2017} suggest lower-bound estimates on the source distance of 6 kpc and 5 kpc, respectively. \citet{Zdziarski2019} determined the M$_{BH}$ of GX 339-4 to be $4-11 M_\odot$ using evolutionary models for the donor. They also suggest the inclination angle ($i$) and the source distance ($D$) to be $40^\circ$--$60^\circ$ and $8-12$ kpc, respectively. In this study, we use averaged values of the M$_{BH}$, $D$, and $i$ for GX 339-4, as reported by \citet{Zdziarski2019}: M$_{BH} = 7.5 M_\odot$, $D$ = 10 kpc, and $i$ = $50^\circ$, as these represent the latest estimates available.

GX 339-4 underwent an X-ray outburst in 2021, which lasted for approximately ten months. Several X-ray missions observed this source during the outburst, and these observations are studied comprehensively by \citet{Wang2021} (NICER; \citealt{Arzoumanian2014, Gendreau2016}), \citet{Garcia2021} (NuSTAR; \citealt{Harrison2013}), and \citet{Liu2021} (Insight-HXMT; \citealt{Zhang2020}). AstroSat \citep{Singh2014} also observed the source during this outburst \citep{Husain2021, 2021ATel14455....1B}. Multiple observations of AstroSat were made, all during the rising phase of the outburst (Figure \ref{fig: MAXI_SWIFT_LC}). The first few observations occurred on February 13, March 02–04, and March 05, 2021, while the last one is an eight-day observation spanning from March 30 to April 06, 2021. \citet{Valentina2022} use the first-day data from the eight-day-long observation to discuss the dual corona comptonization model for type B QPO in GX 339-4. \citet{Mondal2023} and \citet{Jana2024} use the full eight-day data set to discuss the temporal and spectral properties of the source, respectively. \citet{Chand2024} examine the accretion geometry of GX 339-4 in hard states and included data of the observations made on February 13 and March 02, 2021. In a comprehensive spectral and temporal study of the outburst profile of GX 339-4 and H 1743-322, \citet{Aneesha2024} utilize all the data from AstroSat during the 2021 outburst of GX 339-4. 

In this work, we include all the observations of AstroSat taken during the 2021 outburst of the GX 339-4. To understand the behaviour of the QPOs, its energy-dependent properties and its correlation with spectral parameters, we extensively study the spectral and temporal evolution of the source during the SIMS and HIMS. We utilize the scheme discussed in \citet{Garg2020} and \citet{Garg2022} to fit the energy-dependent rms and time-lag of the observed QPO to investigate the possible radiative component responsible for changes in the energy-dependent properties of QPO, using the capabilities of the SXT and LAXPC instruments onboard AstroSat. 

This paper is organized as follows: Section \ref{section: Obs and data reduction} provides details of the AstroSat observations and the data reduction procedures. Section \ref{section : Data Analysis} presents the data analysis, including the light curves, HID, and both spectral (Section \ref{subsec: Spectral Analysis}) and timing analysis (Section \ref{Section : Timing Analysis}). Section \ref{section : Modelling RMS and LAG} discusses the radiative components possibly responsible for the observed QPOs. Finally, Section \ref{section : Discussions and Conclusions} summarizes our key findings and conclusions.

\begin{figure}
    \includegraphics[width=\columnwidth, height=5cm]{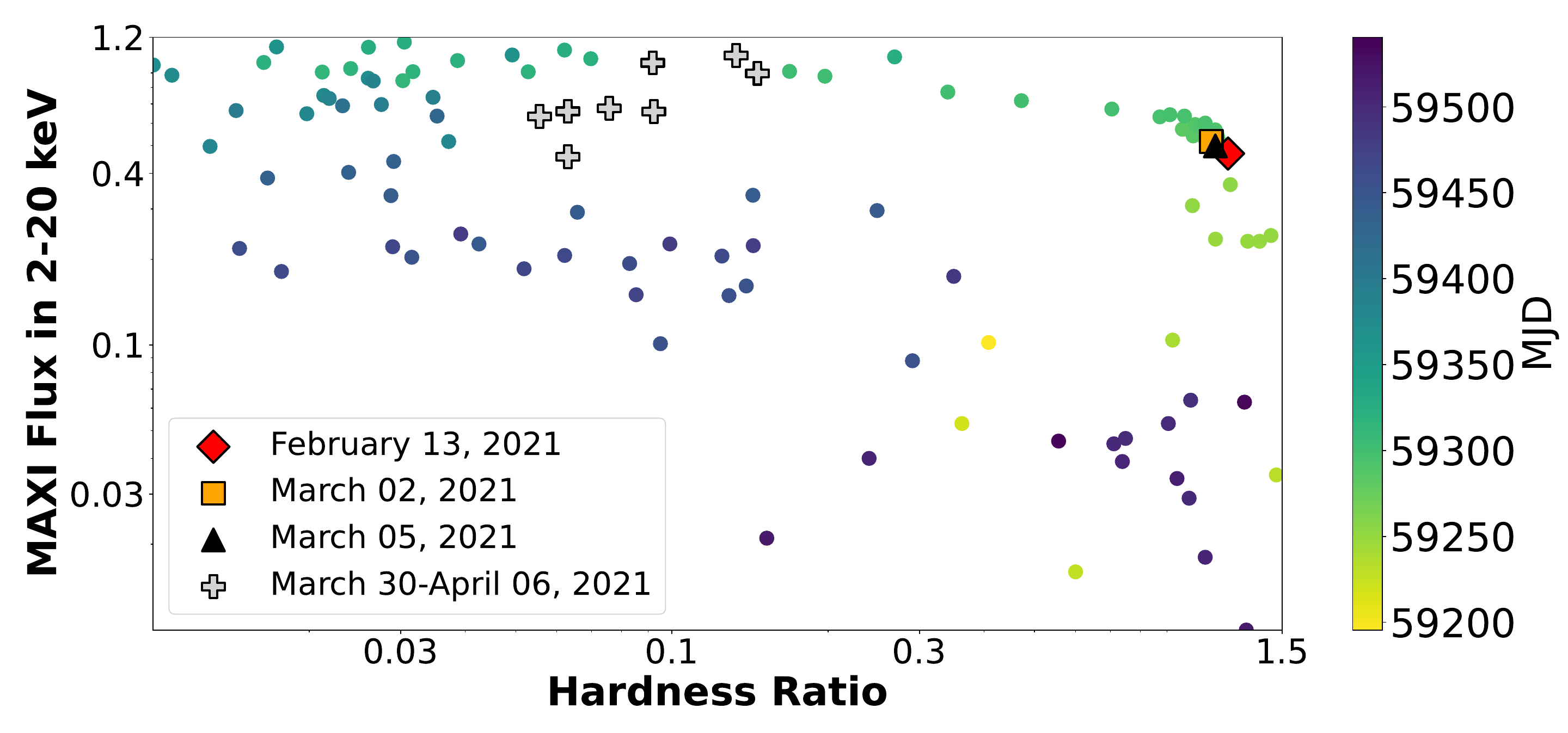}
    \caption{Hardness-Intensity Diagram (HID): The y-axis represents the MAXI flux in the energy range 2-20 keV, while the x-axis shows the HR2, defined as the ratio of Swift/BAT flux in the 15-50 keV range multiplied by 8, divided by the MAXI flux in the 2-20 keV range. The color gradient reflects the time evolution, with the corresponding colorbar displayed on the right side of the figure. Observations O1 (February 13, 2021), O2 (March 02, 2021), and O3 (March 05, 2021) are represented by red diamond, orange square, and black triangle} respectively, while O4$^*$ and O4 (March 30-April 06, 2021) are represented by gray pluses.
    \label{fig: HID}
\end{figure}

\begin{table}
    \centering
    \resizebox{\columnwidth}{!}{ 
    \begin{tabular}{lccrr} 
        \hline \hline
        Observation ID & \makecell{Start Time \\ (MJD)} & \makecell{End Time \\ (MJD)} & \multicolumn{2}{c}{\makecell{Exposure Time \\ (ksec)}} \\
        & & & \makecell{LAXPC} & \makecell{SXT} \\
        \hline
        T03\_275T01\_9000004180$^{O1}$ & 59258.04 & 59258.82 & $\sim$27.04 & $\sim$9.49 \\
        T03\_275T01\_9000004218$^{O2}$ & 59275.04 & 59277.77 & $\sim$99.44 & $\sim$43.23 \\
        T03\_275T01\_9000004222$^{\dagger,O3}$ & 59278.05 & 59278.52 & $\sim$14.56  & $\sim$2.07 \\
        T03\_275T01\_9000004278$^{O4^{*}}$ & 59303.04 & 59304.16 & $\sim$47.65 & $\sim$17.89 \\
        T03\_275T01\_9000004278$^{O4}$ & 59304.17 & 59310.86 & $\sim$251.58 & $\sim$159.81 \\
        \hline
    \end{tabular}
    }
    \caption{AstroSat Observations of GX 339-4 during the 2021 outburst. $\dagger$ indicates the exposure time of three observation IDs: T03\_275T01\_9000004222, T03\_275T01\_9000004224, and T03\_275T01\_9000004226. We merge them in our analysis as these are on the same day. We define the listed observations as O1, O2, O3, O4$^{*}$, and O4 (from top to bottom).}
    \label{tab1}
\end{table}

\section{OBSERVATION AND DATA REDUCTION}
\label{section: Obs and data reduction}

AstroSat, India's first dedicated astronomy satellite, enables simultaneous multi-wavelength observations of astronomical objects with a single satellite. It is equipped with five payloads covering wavelengths from X-ray to UV. One of the significant payloads on AstroSat is the Large Area X-ray Proportional Counters (LAXPC; \citealt{Yadav2016a, Yadav2016b, Agrawal2017, Antia2017}), which consist of three identical but independent proportional counter units (LAXPC10, LAXPC20, and LAXPC30). Each unit has an effective area of 6000 cm$^2$ at 15 keV and a time resolution of 10 µs, operating in the energy range of 3.0–80.0 keV. Another important payload is the Soft X-ray Focusing Telescope (SXT; \citealt{Singh2016, Singh2017}). The SXT has an effective area of 90 cm$^2$ at 1.5 keV and a time resolution of 2.4 seconds (full frame) and 0.278 seconds (150 $\times$ 150 centered pixel frame), covering the energy range of 0.3–8.0 keV.

The HID is shown in figure \ref{fig: HID} using MAXI/GSC (2–20 keV) and Swift/BAT (15–50 keV) flux. We observe that the outburst follows a nearly q-shaped track in the HID. The AstroSat observations are marked in the HID to understand the spectral states of the source during the AstroSat observations.

AstroSat observed the BH LMXB GX 339-4 during its 2021 outburst. The first three observations were conducted on February 13 (O1) \citep{Husain2021}, March 02–04 (O2), and March 05 (O3) \citep{2021ATel14455....1B}, while the last observation spanned eight days from March 30 to April 06, 2021. To understand the spectral and temporal evolution during the eight-day observation, we divide the data sequentially into twenty segments, each with an exposure of about 13–16 ks. We observe QPOs in the first three segments, whereas the remaining seventeen segments do not show any peaked feature in their PDSs. Accordingly, we refer to the first three segments as O4$^*$, where QPOs are present, and the last seventeen segments as O4, where no such feature is observed in the PDS. We further sub-divide the first three segments, now termed as O4$^*$, into fifteen segments, each with an exposure time of $\sim 2500-3000$ seconds, to study the variations in the QPO frequency and its energy-dependent properties over time. The details of all the observations are provided in Table \ref{tab1}.  MAXI flux (2.0–20.0 keV) and normalized Swift/BAT flux (15.0–50.0 keV) are shown in Figure \ref{fig: MAXI_SWIFT_LC} to study the behaviour of soft X-rays and hard X-ray flux during the AstroSat observations. To make the comparison clearer, the Swift/BAT flux is multiplied by 25, allowing its behaviour to be seen alongside the MAXI flux (Figure \ref{fig: MAXI_SWIFT_LC}).

Our spectral analysis use data from AstroSat/LAXPC (4.0–25.0 keV in O4$^{*}$ \& O4 and 5.0-40.0 keV in O1, O2, and O3 and AstroSat/SXT (0.7–5.0 keV in all the observations) after considering the background domination. Using \href{https://www.tifr.res.in/~astrosat_laxpc/LaxpcSoft.html}{LAXPC software\footnote{\url{ https://www.tifr.res.in/~astrosat_laxpc/LaxpcSoft.html}}}, we create a level2.event.fits file from the level1 data of LAXPC20, which we downloaded from the \href{https://astrobrowse.issdc.gov.in/astro_archive/archive/Home.jsp}{AstroSat data archive\footnote{\url{https://astrobrowse.issdc.gov.in/astro_archive/archive/Home.jsp}}} for each observation. Due to low gain and the instrument's response (\citealt{Antia2017, Antia2021}), we do not include data from the LAXPC10 and LAXPC30 PCUs in our study. Our analysis focuses on the spectral and temporal properties derived from the level2 data of LAXPC20 only.

We downloaded the SXT level2 data from the \href{https://astrobrowse.issdc.gov.in/astro_archive/archive/Home.jsp}{AstroSat data archive\footnote{\url{https://astrobrowse.issdc.gov.in/astro_archive/archive/Home.jsp}}}. To create a merged clean event file, we combine the level2 files of all orbits using the \href{https://github.com/gulabd/SXTMerger.jl}{SXT event merger\footnote{\url{https://github.com/gulabd/SXTMerger.jl}}} tool in Julia. This merged file is then used to generate light curves and spectra. In the first three listed observations, there is no need for a pile-up correction. During the last eight-day-long observation, the SXT count rate exceeds 40 counts/sec, which is the suggested threshold for pile-up. To mitigate the pile-up effect, we use an annular region with an inner radius of 7 arcmin and an outer radius of 15 arcmin, using the ds9 software. We use the background file 'SkyBkg\_comb\_EL3p5\_Cl\_Rd16p0\_v01.pha' and the response file 'sxt\_pc\_mat\_g0to12.rmf' provided by the \href{https://www.tifr.res.in/~astrosat_sxt/dataanalysis.html}{AstroSat science support cell\footnote{\url{https://www.tifr.res.in/~astrosat_sxt/dataanalysis.html}}}. The vignetting correction is applied using the \href{https://www.tifr.res.in/~astrosat_sxt/dataanalysis.html}{SXTARFModule\_v02 tool\footnote{\url{https://www.tifr.res.in/~astrosat_sxt/dataanalysis.html}}}. We group the source, response, and background spectrum files using the \texttt{ftgrppha} ftool \citep{Kaastra...2016}.

\begin{figure*}
	\includegraphics[width=\textwidth, height=6.2cm]{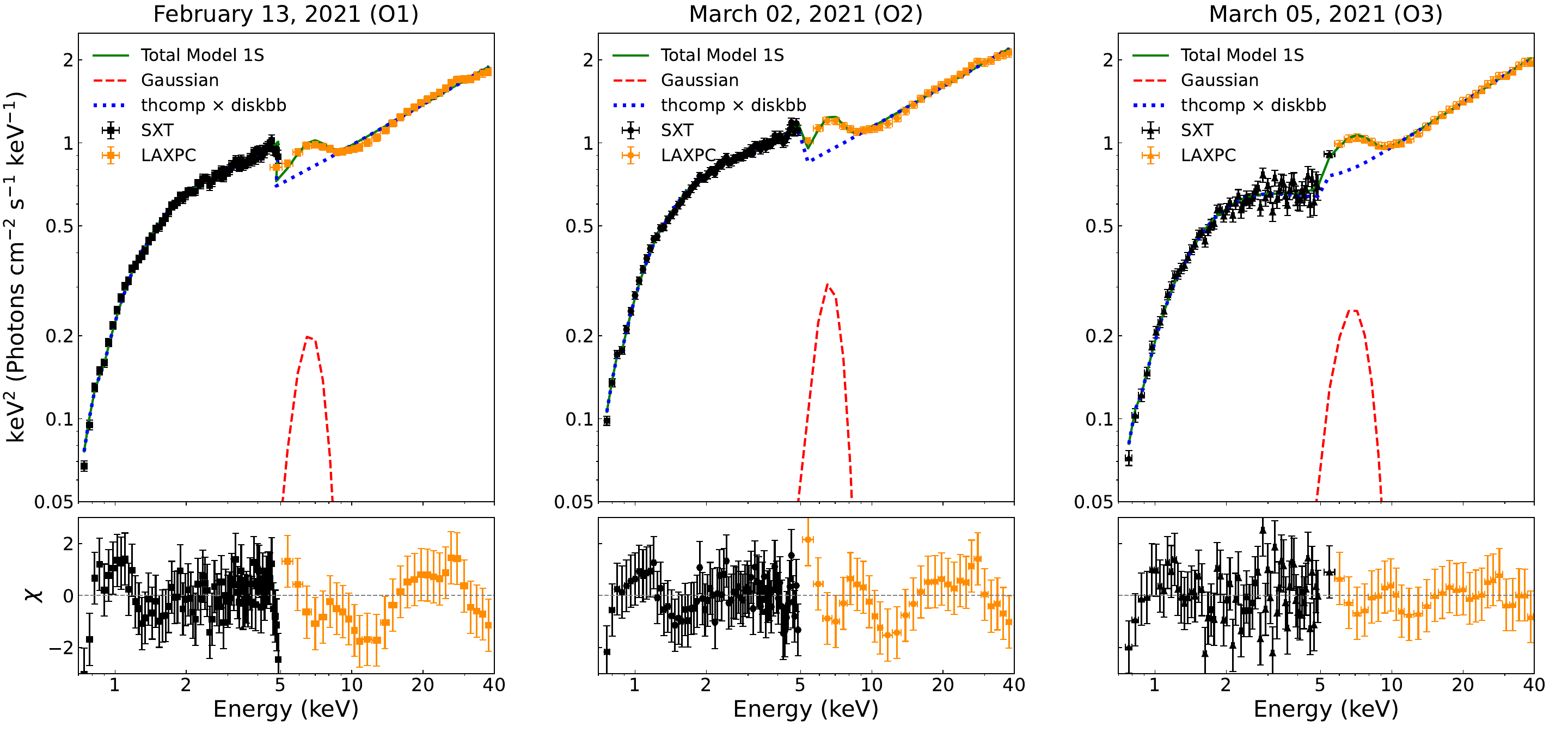}
        \caption{Energy spectrum of observations O1 (Left), O2 (Middle), and O3 (Right), showing SXT (black squares) and LAXPC (orange circles) data points with corresponding error bars. The solid green lines represent the total model fits for each observation, fitted with \textbf{Model 1S}. The bottom panel displays the residuals ($\chi$) for each dataset. Individual model components are displayed: Gaussian (dashed red), and thcomp~$\times$~diskbb (dotted blue).}
    \label{fig: ES_HIMS}
\end{figure*}

\begin{figure*}
	\includegraphics[width=\textwidth, height=6.2cm]{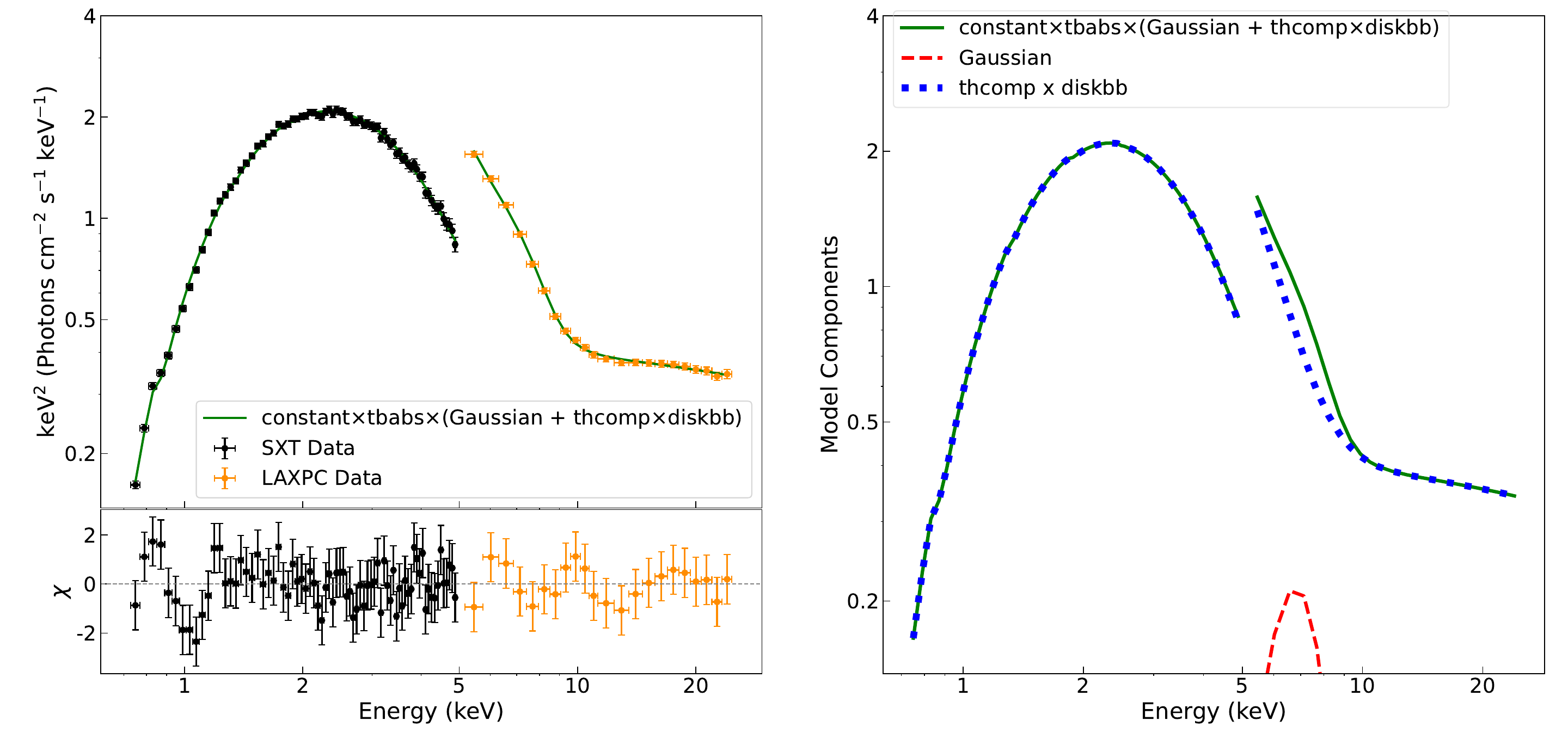}
\caption{\textbf{Energy Spectrum and Model Components}. \textbf{Left Panel:} The upper section shows the observed SXT (black squares) and LAXPC (orange circles) spectra with error bars of segment 1 of O4, overlaid with the best-fit model (solid green line) for \textbf{Model 1S:} \texttt{constant~$\times$~tbabs~$\times$~(gaussian~+~thcomp~$\times$~diskbb)}. The lower section presents the residuals ($\chi$) as a function of energy. \textbf{Right Panel:} Individual model components are displayed: the total model (solid green), Gaussian (dashed red), and thcomp~$\times$~diskbb (dotted blue).}
    \label{fig: ES_diskbb}
\end{figure*}

\begin{figure*}
	\includegraphics[width=0.9\textwidth, height=6.2cm]{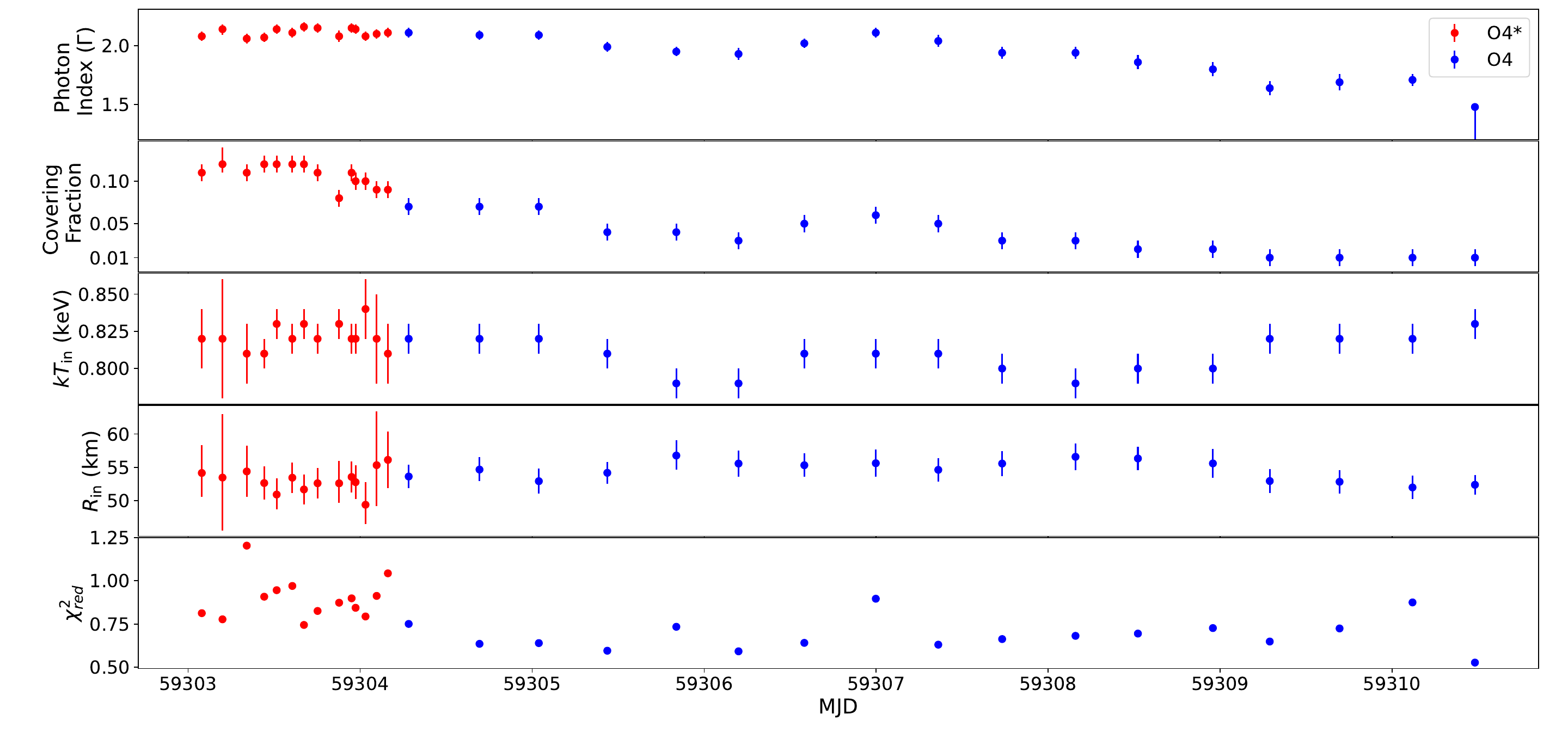}
    \caption{
    Evolution of the spectral parameters and fit statistics for observations O4$^*$ (red) and O4 (blue) as a function of MJD. The panels (top to bottom) show the photon index, covering fraction, inner disc temperature ($kT_{\mathrm{in}}$), apparent inner disc radius, and reduced $\chi^2$ from the spectral fits with \textbf{Model 1S}.}
    \label{fig: Results_Thcomp_Diskbb}
\end{figure*}

\begin{figure*}
	\includegraphics[width=0.9\textwidth, height=6.2cm]{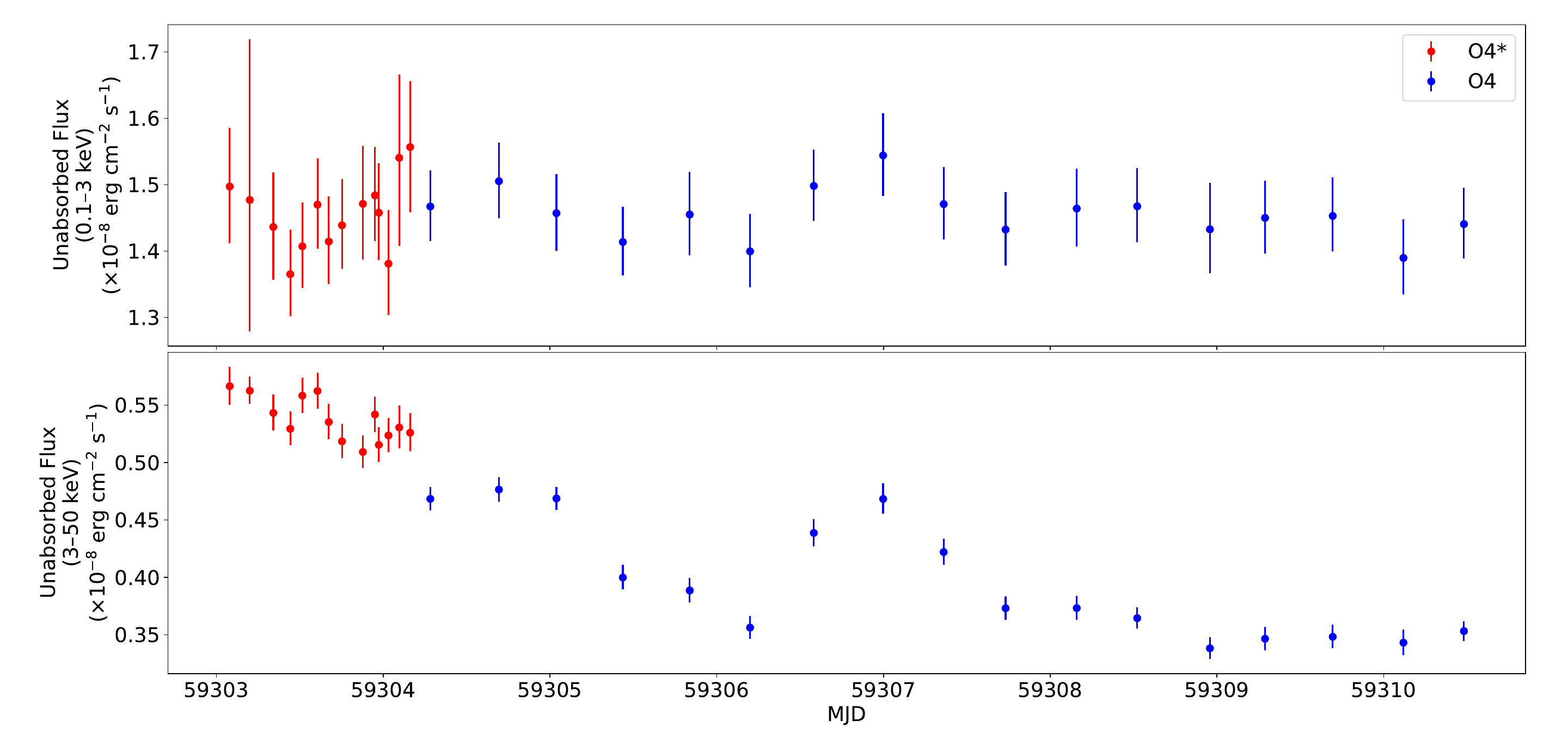}
    \caption{
    Evolution of unabsorbed flux in the energy range 0.1 - 3 keV (upper panel) and 3 - 50 keV (lower panel) for observations O4$^*$ (red) and O4 (blue) as a function of MJD.}
    \label{fig: Unabsorbed Flux}
\end{figure*}

\section{Data Analysis}
\label{section : Data Analysis} 

\subsection{Spectral Analysis}
\label{subsec: Spectral Analysis}

\begin{table}
    \centering
    \resizebox{\columnwidth}{!}{ 
    \renewcommand{\arraystretch}{2}
    
    \begin{tabular}{lcccr}
\hline
Photon Index & Covering Fraction & Inner Disc Temp (keV) & $R_{\mathrm{in}}$ (km) & $\chi^2$/dof \\
\hline
\hline
\multicolumn{5}{c}{\textbf{Observation O1}} \\
\hline
$1.55^{+0.01}_{-0.01}$ & $>0.97$ & $0.53^{+0.06}_{-0.07}$ & $35.52^{+3.06}_{-6.85}$ & 133.54/97\\
\hline
\multicolumn{5}{c}{\textbf{Observation O2}} \\
\hline
$1.56^{+0.01}_{-0.01}$ & $0.95^{+0.03}_{-0.03}$ & $0.48^{+0.05}_{-0.05}$ & $46.84^{+10.43}_{-8.03}$ & 113.25/97 \\
\hline
\multicolumn{5}{c}{\textbf{Observation O3}} \\
\hline
$1.52^{+0.02}_{-0.02}$ & $0.59^{+0.08}_{-0.08}$ & $0.79^{+0.08}_{-0.09}$ & $21.68^{+3.47}_{-2.27}$ & 88.54/88 \\
\hline
\multicolumn{5}{c}{\textbf{Observation O4$^{*}$}} \\
\hline
$2.08^{+0.04}_{-0.04}$ & $0.11^{+0.01}_{-0.01}$ & $0.82^{+0.02}_{-0.02}$ & $54.17^{+4.16}_{-3.58}$ & 63.41/78\\
$2.06^{+0.04}_{-0.04}$ & $0.11^{+0.01}_{-0.01}$ & $0.81^{+0.02}_{-0.02}$ & $54.40^{+3.82}_{-3.32}$ & 96.27/80\\
$2.07^{+0.04}_{-0.04}$ & $0.12^{+0.01}_{-0.01}$ & $0.81^{+0.01}_{-0.01}$ & $52.66^{+2.49}_{-2.26}$ & 77.19/85\\
$2.14^{+0.04}_{-0.04}$ & $0.12^{+0.01}_{-0.01}$ & $0.83^{+0.01}_{-0.01}$ & $50.94^{+2.41}_{-2.19}$ & 80.40/85\\
$2.11^{+0.04}_{-0.04}$ & $0.12^{+0.01}_{-0.01}$ & $0.82^{+0.01}_{-0.01}$ & $53.46^{+2.54}_{-2.30}$ & 82.47/85\\
$2.16^{+0.04}_{-0.04}$ & $0.12^{+0.01}_{-0.01}$ & $0.83^{+0.01}_{-0.01}$ & $51.70^{+2.46}_{-2.24}$ & 63.31/85\\
$2.15^{+0.04}_{-0.04}$ & $0.11^{+0.01}_{-0.01}$ & $0.82^{+0.01}_{-0.01}$ & $52.63^{+2.50}_{-2.28}$ & 70.19/85\\
$2.08^{+0.05}_{-0.05}$ & $0.08^{+0.01}_{-0.01}$ & $0.83^{+0.01}_{-0.01}$ & $52.62^{+3.32}_{-2.93}$ & 72.49/83\\
$2.15^{+0.04}_{-0.04}$ & $0.11^{+0.01}_{-0.01}$ & $0.82^{+0.01}_{-0.01}$ & $53.57^{+2.60}_{-2.36}$ & 76.39/85\\
$2.14^{+0.04}_{-0.04}$ & $0.10^{+0.01}_{-0.01}$ & $0.82^{+0.01}_{-0.01}$ & $52.79^{+2.85}_{-2.55}$ & 71.74/85\\
$2.08^{+0.04}_{-0.04}$ & $0.10^{+0.01}_{-0.01}$ & $0.84^{+0.02}_{-0.02}$ & $49.41^{+3.35}_{-2.92}$ & 64.30/81\\
$2.10^{+0.04}_{-0.04}$ & $0.09^{+0.01}_{-0.01}$ & $0.82^{+0.03}_{-0.03}$ & $55.34^{+8.05}_{-6.09}$ & 66.63/73\\
$2.11^{+0.04}_{-0.04}$ & $0.09^{+0.01}_{-0.01}$ & $0.81^{+0.02}_{-0.02}$ & $56.13^{+5.04}_{-4.21}$ & 79.29/76\\
$2.38^{+0.05}_{-0.05}$ & $0.12^{+0.02}_{-0.02}$ & $0.81^{+0.02}_{-0.02}$ & $48.98^{+4.69}_{-3.89}$ & 71.06/75\\
\hline
\multicolumn{5}{c}{\textbf{Observation O4}} \\
\hline
$2.11^{+0.04}_{-0.04}$ & $0.07^{+0.01}_{-0.01}$ & $0.82^{+0.01}_{-0.01}$ & $53.64^{+1.73}_{-1.62}$ & 67.57/90\\
$2.09^{+0.04}_{-0.04}$ & $0.07^{+0.01}_{-0.01}$ & $0.82^{+0.01}_{-0.01}$ & $54.68^{+1.83}_{-1.71}$ & 56.57/89\\
$2.09^{+0.04}_{-0.04}$ & $0.07^{+0.01}_{-0.01}$ & $0.82^{+0.01}_{-0.01}$ & $52.94^{+2.07}_{-1.89}$ & 56.31/88\\
$1.99^{+0.04}_{-0.04}$ & $0.04^{+0.01}_{-0.01}$ & $0.81^{+0.01}_{-0.01}$ & $54.19^{+1.71}_{-1.61}$ & 54.22/91\\
$1.95^{+0.04}_{-0.04}$ & $0.04^{+0.01}_{-0.01}$ & $0.79^{+0.01}_{-0.01}$ & $56.78^{+2.29}_{-2.09}$ & 63.87/87\\
$1.93^{+0.05}_{-0.05}$ & $0.03^{+0.01}_{-0.01}$ & $0.79^{+0.01}_{-0.01}$ & $55.58^{+1.95}_{-1.83}$ & 52.72/89\\
$2.02^{+0.04}_{-0.04}$ & $0.05^{+0.01}_{-0.01}$ & $0.81^{+0.01}_{-0.01}$ & $55.33^{+1.75}_{-1.64}$ & 58.35/91\\
$2.11^{+0.04}_{-0.04}$ & $0.06^{+0.01}_{-0.01}$ & $0.81^{+0.01}_{-0.01}$ & $55.63^{+2.21}_{-2.04}$ & 77.99/87\\
$2.04^{+0.05}_{-0.05}$ & $0.05^{+0.01}_{-0.01}$ & $0.81^{+0.01}_{-0.01}$ & $54.64^{+1.77}_{-1.67}$ & 56.77/90\\
$1.94^{+0.05}_{-0.05}$ & $0.03^{+0.01}_{-0.01}$ & $0.80^{+0.01}_{-0.01}$ & $55.58^{+1.85}_{-1.74}$ & 59.02/89\\
$1.94^{+0.05}_{-0.05}$ & $0.03^{+0.01}_{-0.01}$ & $0.79^{+0.01}_{-0.01}$ & $56.59^{+2.03}_{-1.90}$ & 60.00/88\\
$1.86^{+0.06}_{-0.06}$ & $0.02^{+0.01}_{-0.01}$ & $0.80^{+0.01}_{-0.01}$ & $56.34^{+1.76}_{-1.67}$ & 63.94/92\\
$1.80^{+0.06}_{-0.06}$ & $0.02^{+0.01}_{-0.01}$ & $0.80^{+0.01}_{-0.01}$ & $55.60^{+2.31}_{-2.13}$ & 63.23/87\\
$1.64^{+0.06}_{-0.06}$ & $0.01^{+0.01}_{-0.01}$ & $0.82^{+0.01}_{-0.01}$ & $52.96^{+1.76}_{-1.63}$ & 59.05/91\\
$1.69^{+0.06}_{-0.07}$ & $0.01^{+0.01}_{-0.01}$ & $0.82^{+0.01}_{-0.01}$ & $52.85^{+1.74}_{-1.61}$ & 65.94/91\\
$1.71^{+0.05}_{-0.05}$ & $0.01^{+0.01}_{-0.01}$ & $0.82^{+0.01}_{-0.01}$ & $52.00^{+1.91}_{-1.77}$ & 77.88/89\\
\hline
\end{tabular}
    }
    \caption{Best-fit spectral parameters for all observations fitted with \textbf{Model 1S}. This table shows the values of photon index, covering fraction, inner disc temperature, and the "apparent" inner disc radius ($R_{\mathrm{in}}$) in kilometers with $90\%$ confidence range}.
    \label{Table: Model1S Parameters}
\end{table}
To study the spectral evolution during the presence of QPOs, we subdivide O4$^{*}$ into fifteen segments (see Section \ref{section: Obs and data reduction} for segmentation details). Similarly, we divide O4 into seventeen segments (see Section \ref{section: Obs and data reduction} for segmentation details) to understand the evolution of the outburst in the absence of QPOs. We perform the spectral analysis on O1, O2, O3, and on the total thirty-two segments of observations O4$^{*}$ and O4.
 
We begin our spectral analysis with a multi-colour disc blackbody (\texttt{diskbb}; \citealt{mitsuda1984}) to model the thermal emission from the accretion disc, and \texttt{thcomp} \citep{Zdziarski2020} to fit the non-thermal emission. We modify the total model by multiplying a Galactic absorption component \texttt{tbabs} \citep{Wilms2000}, using \texttt{abund wilm}, as recommended in the XSPEC manual. To account for the differences in the normalization between the SXT and LAXPC instruments, we multiply the total model by a \texttt{constant} factor. The constant factor for LAXPC is fixed at 1, while for SXT, it is kept free. We apply systematic errors of 2\% to account for uncertainties in the instrument calibration. We also use the \texttt{gain fit} command within XSPEC to correct for the gain shift in the SXT data during spectral fitting and use the \texttt{"energies 0.1 300 1000 log"} command when applying the \textbf{thcomp} model. After fitting the energy spectrum with the above model, residuals indicate that additional fitting, specifically a \texttt{gaussian} component to account for strong residuals around 6.4 keV, is necessary. Therefore, we introduce a \texttt{gaussian} at 6.4 keV.

Thus, the model become \textit{\textbf{constant $\times$ tbabs $\times$ (gaussian + thcomp $\times$ diskbb)}}, refer to as \textbf{Model 1S}. We apply \textbf{Model 1S} to all the observations, and we get acceptable fit as we get $\chi^{2}_{red}$ to be 1.03, 0.63, and 0.82 in O1, O2, and O3 respectively (Figure \ref{fig: ES_HIMS}) while $\chi^{2}_{red}$ lies between 0.74-1.02 in O4$^{*}$ and 0.53-0.90 in O4. The energy spectrum fitting for one of the segments of observation O4 is shown in Figure \ref{fig: ES_diskbb}. Offset values after applying the \texttt{gain fit} for the energy spectra fitted with \textbf{Model 1S} are $-0.73$ eV, $7.38$ eV, and $-13.8$ eV for observations O1, O2, and O3, respectively. For observation O4$^*$, the offset values range from $5.69$ eV to $19.7$ eV, while for observation O4, they range from $8.29$ eV to $17.2$ eV. The best-fit spectral parameters such as photon index, covering fraction, inner disc temperature, and apparent inner disc radius, of all the observations are listed in the Table \ref{Table: Model1S Parameters}. Figure \ref{fig: Results_Thcomp_Diskbb} shows the variation of the spectral parameters during observations O4$^*$ and O4. We calculated apparent inner disc radius using the relation \texttt{diskbb} normalization ($N_{dbb}$) = $(R_{in}/D_{10})^2cos\theta$, where $R_{in}$ is the "apparent" inner disc radius in km, $D_{10}$ is the source distance in units of 10 kpc, and $\theta$ is the inclination angle of the disc. We use source distance to be 10 kpc and inclination angle of the disc to be 50$^\circ$ in the calculation of the apparent inner disc radius \citep{Zdziarski2019}. Further, we studied the evolution of unabsorbed flux in the energy range 0.1-3 keV (soft) and 3-50 keV (hard) which is given in Figure \ref{fig: Unabsorbed Flux}. The unabsorbed flux is stable in the softer energy range while it decreases in the harder energy range.

\subsection{Timing Analysis}
\label{Section : Timing Analysis}

We generate PDSs for each observation using \texttt{laxpc\_find\_freqlag}, a routine available in the LAXPCSoftware. The PDSs created using this routine are rms normalized, dead time Poisson level and background corrected. We use a time resolution of $\sim 6.25$ ms and 16,384 bins per segment, resulting in a total segment length of $\sim 102.4$ s, with the minimum and Nyquist frequencies being $\sim 0.01$ Hz and $\sim 80$ Hz, respectively to generate the PDSs in each segment. We observe a peak-like structure in O1, O2, O3, and O4$^{*}$, but no such feature is observed in O4. In O4$^{*}$, the PDS shows a broad peak, suggesting a gradual evolution of the QPO centroid frequency over time. To examine the time evolution of the QPO, we divide O4$^{*}$ into fifteen segments of approximately equal duration ($\sim 2500-3000$ seconds) as discussed in the Section \ref{section: Obs and data reduction} and generate the PDS for each segment. We also analyze the peak-like feature in the PDS of O1, O2, and O3 but find no significant variation in QPO frequency within the observations. To determine the centroid frequency of the QPOs, we fit the PDSs with the \texttt{Lorentzian} model within \texttt{XSPEC}.

\begin{figure}
	\includegraphics[width=\columnwidth, height=5.2cm]{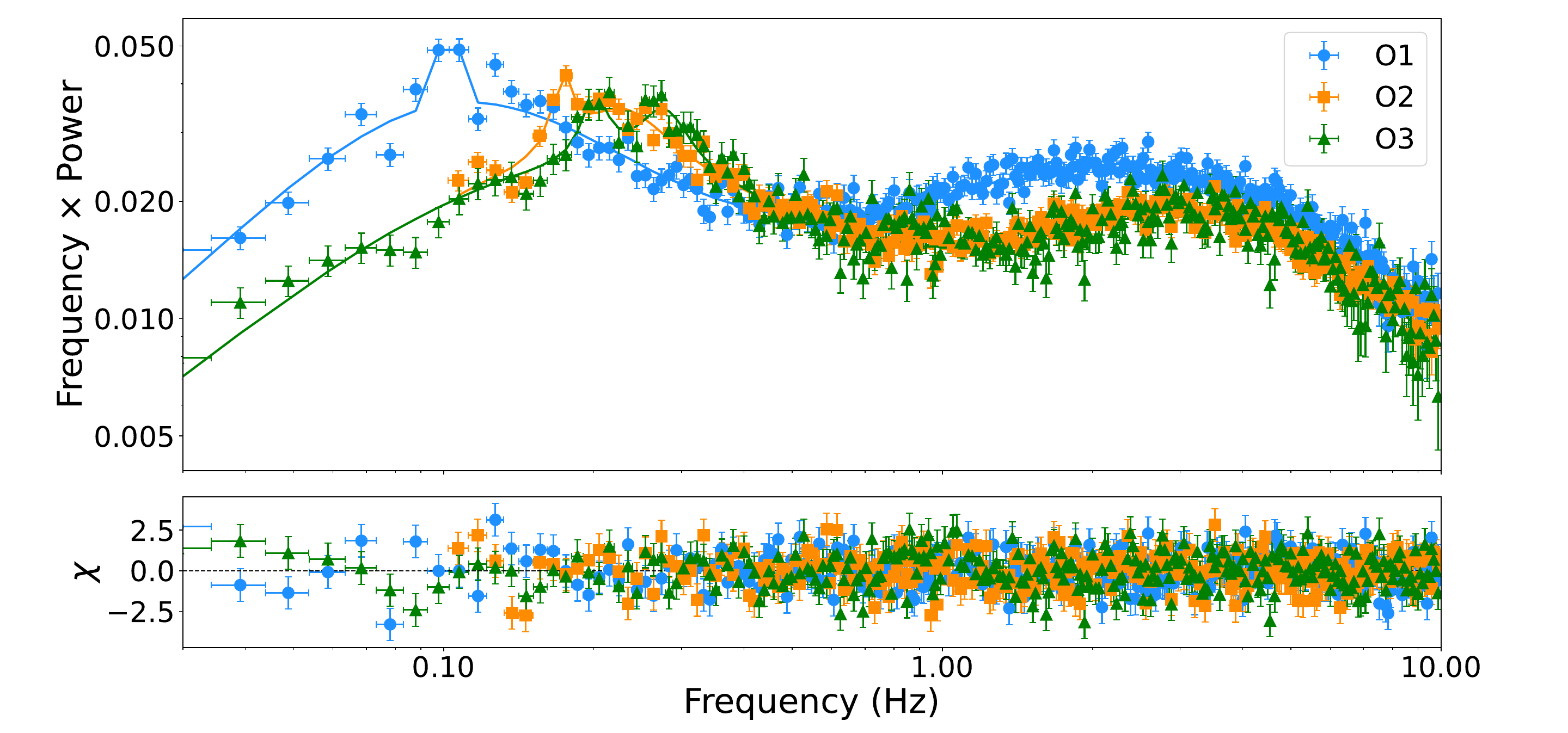}
    \caption{Power density spectra (PDS) for the observations O1, O2, and O3. 
    The top panel shows frequency times power ($f \times P(f)$) vs. frequency (Hz), with \textcolor{blue}{blue circles} (O1), \textcolor{orange}{orange squares} (O2), and \textcolor{green}{green triangles} (O3). Solid lines represent model fits. Power in the PDSs are rms normalised.} The bottom panel displays residuals $(\text{Data} - \text{Model}) / \text{Error}$.

    \label{fig: PDS_O1_O2_O3}
\end{figure}

During the O1, O2, O3, we observe that the QPO frequency increases from $0.10^{+0.01}_{-0.01}$ Hz in O1 to $0.18^{+0.01}_{-0.01}$ Hz in O2, and further to $0.20^{+0.01}_{-0.01}$ Hz in O3 (Figure \ref{fig: PDS_O1_O2_O3}) which is also reported by \cite{Aneesha2024}, and \cite{Chand2024}. No harmonics or subharmonics are observed during these observations. We observe that the QPOs in these observations have frequencies $\sim0.1-0.2$ Hz, quality factor (Q) $\geq 6$, and rms $\geq 3\%$ hence we can classify these QPOs as type C \citep{Casella2005,Ingram2019}.

\begin{figure}
    \includegraphics[width=\columnwidth, height=5.2cm]{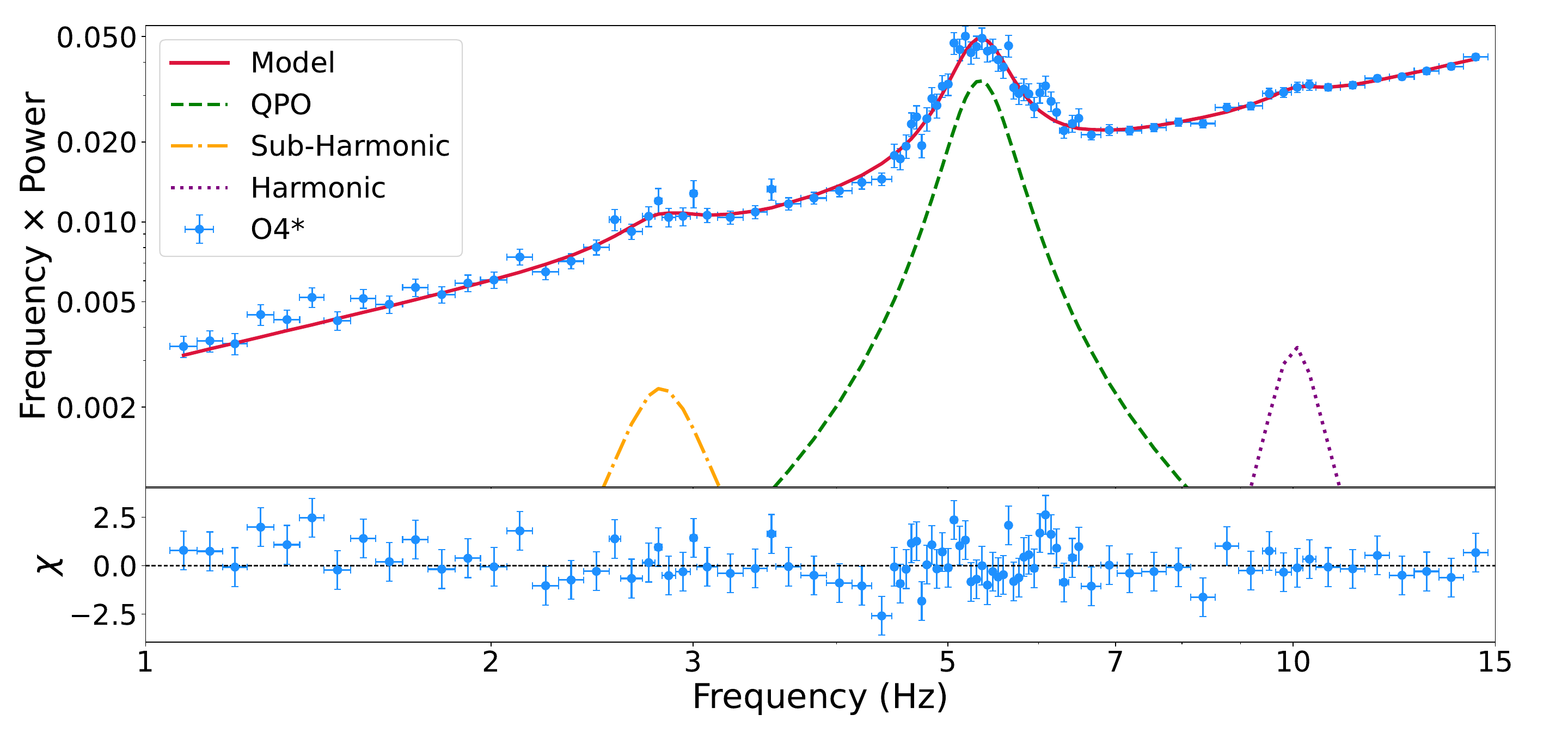}
    \caption{Power density spectrum (PDS) of the observation O4$^{*}$. 
    The \textbf{upper panel} shows the PDS, fitted with multiple \textbf{Lorentzian} components and a \textbf{power-law}. Blue points represent frequency $\times$ power, while the solid red line indicates the best-fit model. Power in the PDS is rms normalised. 
    The \textbf{lower panel} displays the residuals ($\chi$ values) as green points.}
    \label{fig: PDS_seg2_O4*}
\end{figure}

\begin{figure}
    \includegraphics[width=\columnwidth, height=5cm]{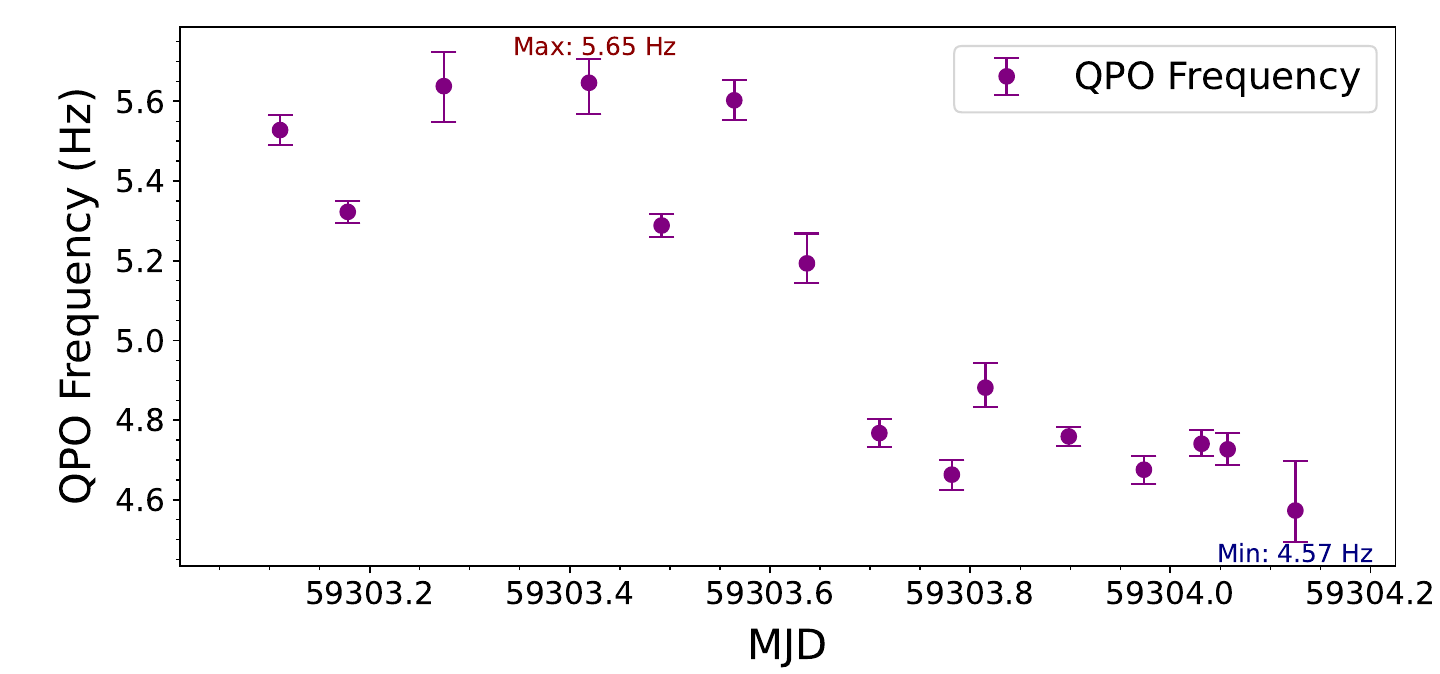}
    \caption{QPO Evolution: The y-axis represents the QPO frequency, while the x-axis shows the modified julian date (MJD). The QPO frequency evolves from approximately 5.7 Hz to 4.5 Hz over time.}
    \label{fig: QPOvariation}
\end{figure}

In the O4$^{*}$, harmonics and subharmonics are observed in some segments, which have been reported in the earlier studies by \citet{Aneesha2024} and \citet{Mondal2023}. For these PDS fits, we initially use three Lorentzians, along with a power-law component where the photon index is fixed at 0. The number of \texttt{Lorentzians} is adjusted as needed to account for the presence of harmonics and subharmonics. Figure \ref{fig: PDS_seg2_O4*} represents the PDS of seg 2 of O4$^{*}$ fitted with multiple \texttt{Lorentzians} and \texttt{powerlaw}. After fitting PDSs of all the segments of O4$^{*}$, we observe a clear evolution of the QPO frequency, which decreases from $5.65^{+0.06}_{-0.08}$ Hz to $4.57^{+0.12}_{-0.08}$ Hz over time (Figure \ref{fig: QPOvariation}). In this observation, the QPOs show centroid frequencies of approximately 4.5–5.5 Hz, quality factors $\geq 6$ (within 3$\sigma$), and rms amplitudes of about 4–8$\%$. Based on the classification criteria discussed by \citet{Casella2005} and \citet{Ingram2009}, these QPOs can be best classified as type-B. \citet{Valentina2022} also describe these QPOs as type B in their study using same AstroSat data. \citet{Aneesha2024} also reported QPO frequency $5.37^{+0.06}_{-0.02}$ Hz along with harmonics and subharmonics at $9.38^{+0.33}_{-0.37}$ Hz and $2.86^{+0.08}_{-0.11}$ Hz respectively during the initial phase of O4$^*$ and find a decrease in the QPO frequency later in time within this observation. However, they did not study the evolution of the QPO frequency over time with proper segmentation. \citet{Mondal2023} performed the orbit-wise PDSs fitting and reported the decrease in QPO frequency. They also reported harmonics and subharmonics in some of the orbits, not in all.

\begin{figure*}

	\includegraphics[width=\textwidth]{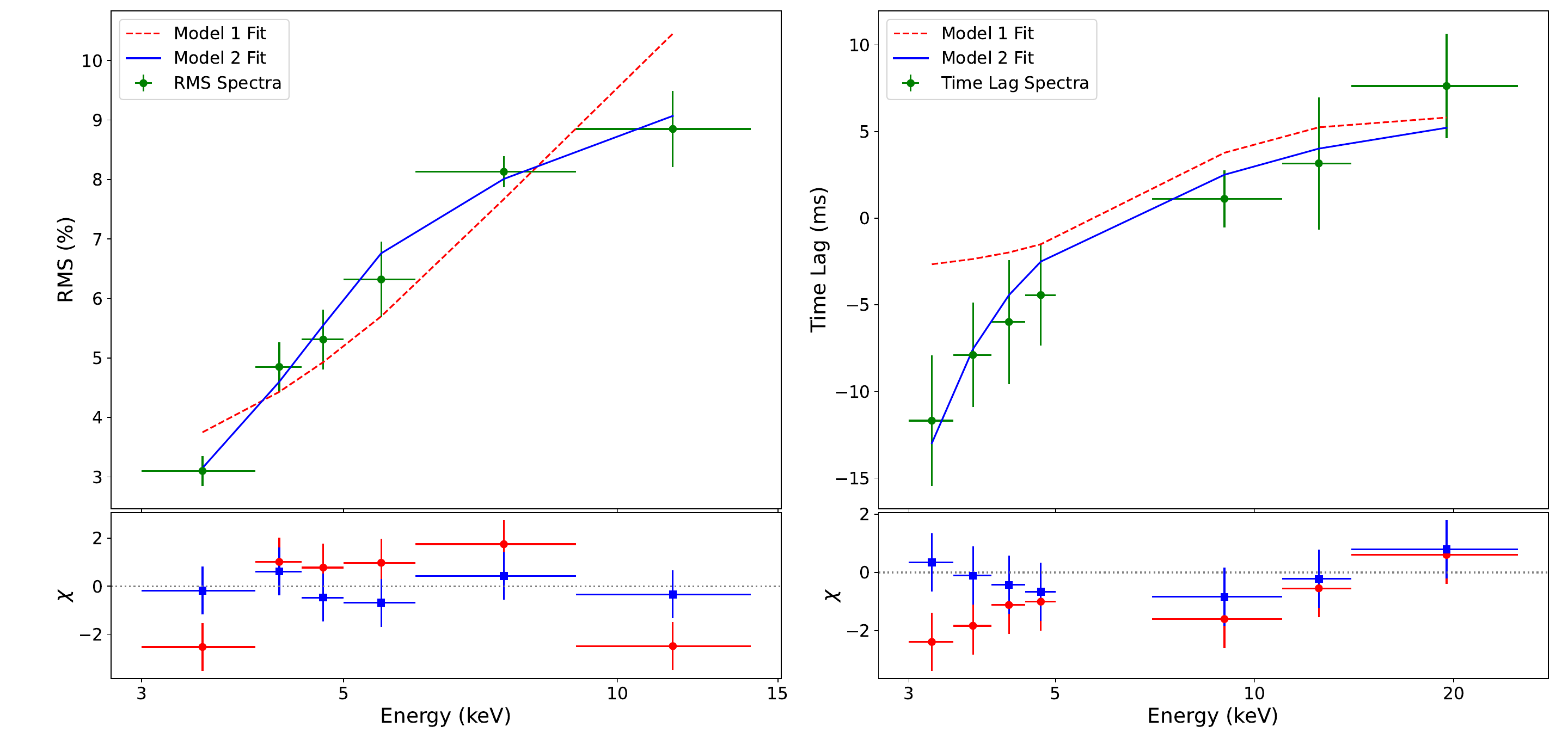}
    \caption{The RMS and time-lag spectra fitted with \textbf{Model 1T} and \textbf{Model 2T} discussed in Section \ref{section : Modelling RMS and LAG}. Green data points in the upper left panel represent the rms in percentage while that in upper right panel represent the time lag (ms). Dashed red and blue line represent the model 1T and model 2T fitting. The lower panel of both left and right panel display the $\chi$ corresponding to model 1T and model 2T fitting with red and blue data points, respectively. We use the $5-7$ keV energy band as a reference to calculate the time lag spectra.}
    \label{fig: LAG_RMS_Model_Fitting}
\end{figure*}

Further, we study the energy-dependent rms and time lag of the QPO during O4$^{*}$ in the energy range of 3-30 keV. For this purpose, we use a subroutine provided by the AstroSat team, \texttt{laxpc$\_$find$\_$freqlag}, which takes characteristic frequency features ($\nu_{0}$) and frequency resolution ($-l$) as input, along with energy bands and GTI, to compute time lags and rms. Typically, we provide the QPO frequency {as $\nu_0$ and $-l$ depends on the total length of the time series.} We use the 5.0-7.0 keV energy band as a reference for estimating the time lags (Figure \ref{fig: LAG_RMS_Model_Fitting}).

We observe that the time lag and rms increase in most of the segments with energy, which has been reported by \citet{Mondal2023} in their orbit-wise analysis. 

\section{Modelling RMS and LAG}
\label{section : Modelling RMS and LAG}
\cite{Garg2020} discuss a generic scheme to model the energy-dependent rms and time lags to identify the radiative components responsible for the QPOs observed in XRBs. In this scheme, they assumed that the energy spectrum of BHXBs consists of two components: a thermal and a non-thermal component, which can be modeled by \texttt{diskbb} and \texttt{nthcomp} within \texttt{XSPEC}. They prefer using physical and geometrical parameters over phenomenological spectral parameters. For example, they convert the coronal electron temperature (kT$_{e}$), a spectral parameter of \texttt{nthcomp}, into the coronal heating rate ($\dot{H}$).
 Specifically, following the formulation of \citet{Garg2020}, $\dot{H}$ is evaluated using the integral:
\begin{equation}
\dot{H} = \int E \left[ f \cdot F_c(E, kT_e, \tau) - f \cdot F_d(E, kT_{\mathrm{in}}) \right] dE, \tag{1} \label{eq:Hdot}
\end{equation}
where $F_c(E)$ is the photon flux from the corona, $F_d(E)$ is the disc flux, and $f$ is the covering fraction from the \texttt{Thcomp} model, representing the disk photons undergoing Comptonization in the corona. The integrand in Equation~\ref{eq:Hdot} multiplies the photon flux difference by photon energy $E$, thereby converting the photon flux into an energy flux. Equation \ref{eq:Hdot} is used to calculate the $\dot{H}$ corresponding to the best-fit spectral parameters obtained from the time-averaged spectral fitting, including the $kT_e$, $kT_{in}$, and optical depth ($\tau$).

\begin{figure}
    \includegraphics[width=\columnwidth]{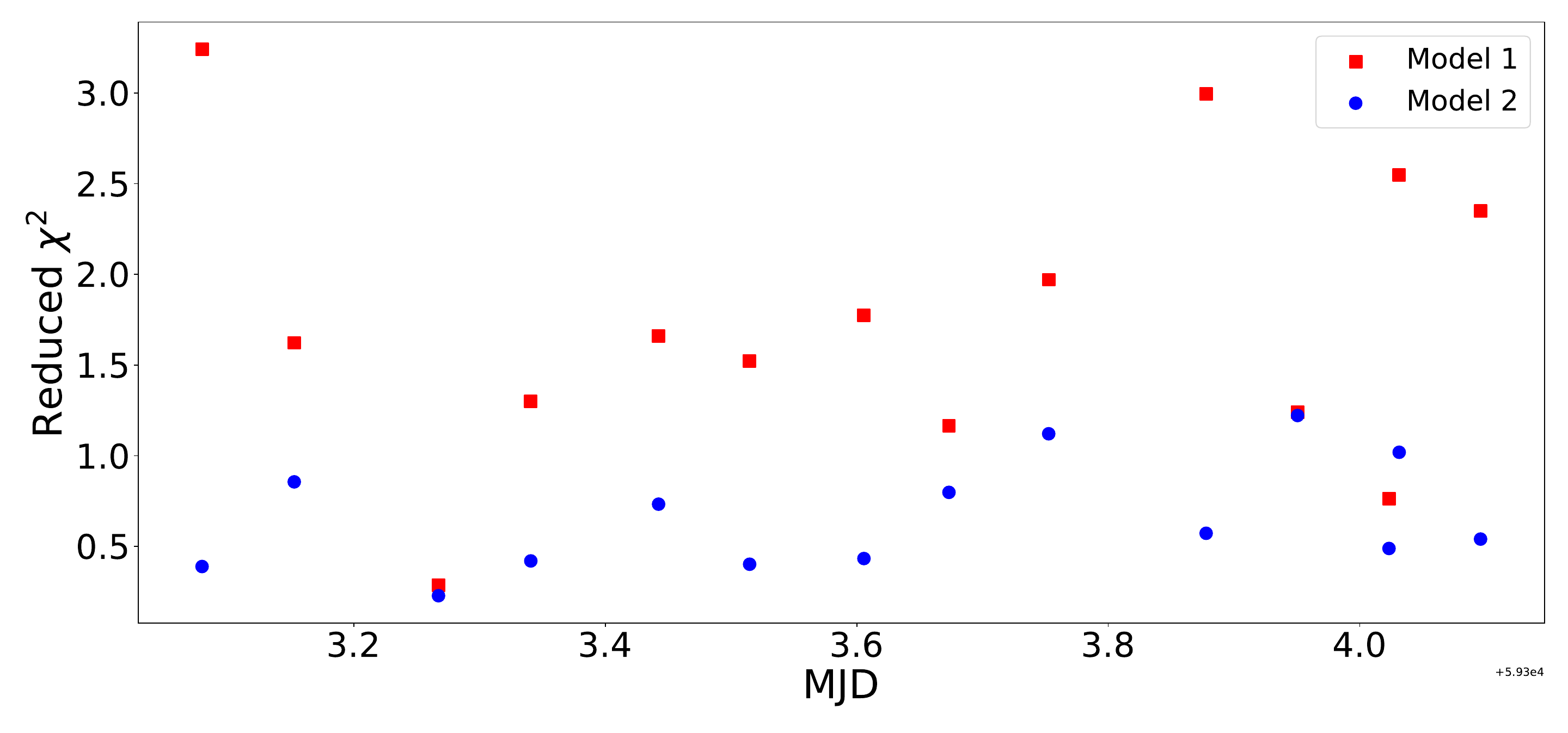}
    \caption{Reduced $\chi^2$ observed in different segments after fitting the rms and time-lag spectra with \textbf{Model 1T} (Red) and \textbf{Model 2T} (Blue)}
    \label{fig: Chi2_comparision_M1_M2}
\end{figure}

\begin{figure*}
	\includegraphics[width=\textwidth]{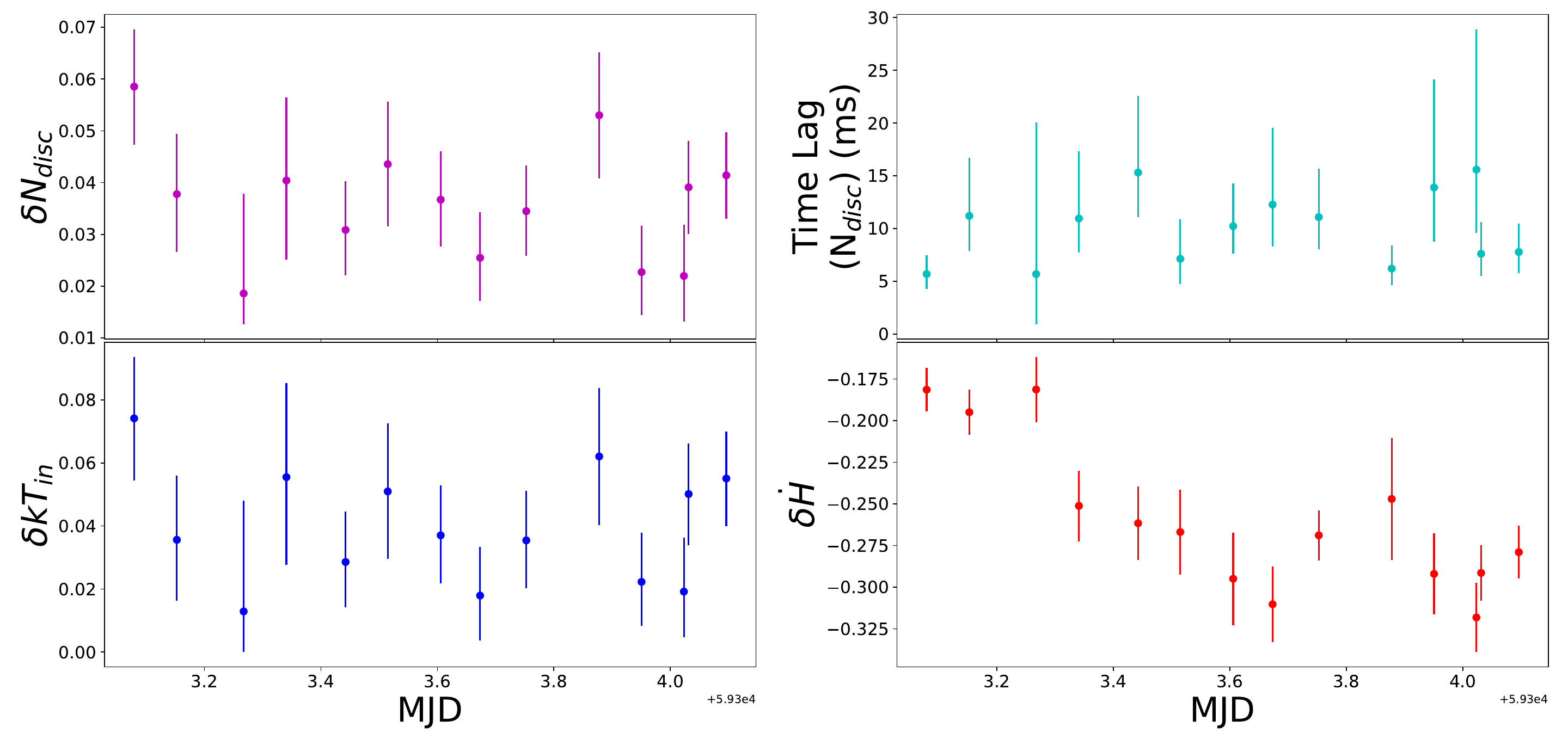}
        \caption{Observed variations in different spectral parameters after fitting the rms and time lag spectra with \textbf{Model 2T}. The observed variations in disc normalization, and inner disc temperature in different segments are shown in the upper, and lower panels of the left column, respectively. The time lag between disc normalization and disc temperature, and the observed variation in heating rate are shown in the upper, and lower panels of the right column, respectively.}

    \label{fig: Model_2_results_frms_lag}
\end{figure*}

In the modeling of the energy-dependent rms and time lag spectra, $\dot{H}$ is treated as the variable physical parameter. For each input value of $\dot{H}$, the model iteratively adjusts $kT_e$ until the computed Comptonized spectrum satisfies Equation~\ref{eq:Hdot}. In this way, the $kT_e$ is used to calculate the $\dot{H}$.
This scheme calculates the \texttt{first-order variation} in the photon spectrum caused by small changes in the physical parameters using the equations discussed in \citet{Garg2020}.

We use the methodology discussed in \cite{Garg2022} to model the energy-dependent time lags and rms. \citet{Garg2022} modify the model discussed in \cite{Garg2020} and replace \texttt{nthcomp} with \texttt{thcomp} to fit the energy spectrum. We fit the energy spectrum of all the segments with the XSPEC model \textbf{constant $\times$ tbabs $\times$ (gaussian + thcomp $\times$ diskbb)} as discussed in Section \ref{subsec: Spectral Analysis}. As described in \citet{Garg2022}, we freeze all the spectral parameters to the best-fit values and replaced \texttt{thcomp} with the \texttt{thcompph} model, which converts the kT$_{e}$ into the physical parameter, $\dot H$. We keep $\dot H$ free and fit the spectrum to obtain its best-fit value.

After fitting the energy spectra of all the segments with the model \textbf{constant $\times$ tbabs $\times$ (gaussian + thcompph $\times$ diskbb)}, as prescribed by the scheme, we proceed to fit the time lag and rms spectra using the model discussed in \citet{Garg2020} and \citet{Garg2022}. We begin by considering variations in the disc temperature ($|\delta$kT$_{in}|$) and heating rate ($|\delta\dot H|$), with a phase-lag $\phi_{\dot H}$ between them. We refer to this as \textbf{Model 1T} (Figure \ref{fig: LAG_RMS_Model_Fitting}). However, we are unable achieve a good fit for most of the segments, as we observe reduced $\chi^{2} > 2$ ({Figure \ref{fig: Chi2_comparision_M1_M2})

To improve the fit, we introduce one more variation in terms of disc normalization ($|\delta$ N$_{disc}|$) and a phase-lag $\phi_{N_{disc}}$ between $|\delta$ N$_{disc}|$ and $|\delta$kT$_{in}|$. This model is refer to as \textbf{Model 2T}. With this model fitting, we observe that phase-lag $\phi_{\dot H}$ is zero between $|\delta\dot H|$ and $|\delta$kT$_{in}|$, hence we freeze it at zero.} The introduction of $|\delta$ N$_{disc}|$ improves the model fit for most segments (Figure \ref{fig: Model_2_results_frms_lag}), with $\chi^{2}_{red}$ found to be less than 1 in most of the segments (Figure \ref{fig: Chi2_comparision_M1_M2}).

\section{Discussions and Conclusions}
\label{section : Discussions and Conclusions}

We study the spectral and temporal properties of the BH LMXB GX 339-4 using \textit{AstroSat}'s LAXPC and SXT data during its 2021 outburst. After dividing the data into multiple segments as discussed in Section \ref{subsec: Spectral Analysis}, we fit the energy spectrum with Model 1S: \textbf{constant $\times$ tbabs $\times$ (gaussian + thcomp $\times$ diskbb)}, assuming that the energy spectrum of the BH LMXB consists primarily of two components: a thermal component and a non-thermal component. This model provided a good fit (Figure \ref{fig: ES_HIMS} $\&$ \ref{fig: ES_diskbb}).

We observe the photon index to be $1.55^{+0.01}_{-0.01}$, $1.56^{+0.01}_{-0.01}$, and $1.52^{+0.02}_{-0.02}$ during the observations O1, O2, and O3, respectively, indicating a relatively hard spectral state in these observations. In contrast, during the observation O4$^*$, the photon index remained steady around $\sim2.11$, suggesting a transition to a softer state. However, during the observation O4, the photon index decreased from $2.11^{+0.04}{-0.04}$ to $1.71^{+0.05}_{-0.05}$ (Figure \ref{fig: Results_Thcomp_Diskbb}), pointing toward spectral hardening during this time. The covering fraction also showed significant changes across observations. It is quite high during O1 and O2, with values $>0.93$ and $0.95^{+0.04}_{-0.04}$ respectively, and moderately lower in O3 at $0.62^{+0.08}{-0.08}$. In contrast, the covering fraction dropped to very low values during both O4$^*$ (below 0.14) and O4 (below 0.08), with a clear declining trend across O4 (Table \ref{Table: Model1S Parameters} $\&$ Figure \ref{fig: Results_Thcomp_Diskbb}).

The inner disc temperature vary between $0.46$ and $0.87$ keV during the observations O1–O3, but remain stable around $\sim0.82$ keV throughout both O4$^*$ and O4. Interestingly, we find that the apparent inner disc radius to be lower during O1–O3 (HIMS) compared to O4$^*$ and O4, which is somewhat unexpected. One possible explanation for this discrepancy could be a change in the color correction factor between these states. To explore this further, we added an additional warm Comptonization component (\texttt{thcomp}) to our base model (Model 1S). With this updated model, we found that the upper bound of the \texttt{diskbb} normalization become unconstrained, which supports the idea that the true inner disc radius might be larger than previously estimated. In O4$^*$ and O4, the inner radius remained stable at around $\sim55$ km, showing small variation within the uncertainties. These results are also reflected in the broadband flux behavior. As shown in Figure \ref{fig: Unabsorbed Flux}, the soft X-ray flux (0.1–3 keV) remains mostly constant, consistent with the small variation in disc parameters (e.g., $kT_{\rm in}$ and $R_{\rm in}$). On the other hand, the hard X-ray flux (3–50 keV) declines with time, likely driving the changes observed in the HID during O4$^*$ and O4 (Figure \ref{fig: HID}).

Furthermore, we observe the QPOs in all the observations except O4. We fit the PDS with multiple \textbf{Lorentzians} along with a \textbf{power law} component, keeping the photon index fixed at $0$. We observe harmonics and subharmonics in some segments of O4$^{*}$, but not in all. During O1, O2, and O3, we observe a steady increase in the QPO frequency, rising from 0.10 Hz in O1 to 0.18 Hz in O2, and further to 0.20 Hz in O3, which is earlier reported by \citet{Aneesha2024}, and \citet{Chand2024}. Notably, no harmonics or subharmonics are detected during these observations. In contrast, during O4$^{*}$, harmonics and subharmonics are observed in some segments. A clear evolution in the QPO frequency is observed in this observation, with the frequency gradually decreasing from $\sim5.65$ Hz to $\sim4.57$ Hz over time (Figure \ref{fig: QPOvariation}). \citet{Mondal2023} also reported the gradual decrease in the QPO frequency in their orbit-wise analysis, as well as harmonics and subharmonics in a few orbits. Our spectral and temporal analysis suggest that the source remains in the HIMS during the first three observations (O1, O2, and O3), and subsequently transitions to the SIMS during the eight-day observation. Specifically, in observations O1, O2, and O3, we measure a photon index of $\sim 1.5$ (Table \ref{Table: Model1S Parameters}) along with the presence of type-C QPOs (Figure \ref{fig: PDS_O1_O2_O3}), which are the characteristic of the HIMS \citep{HomanBelloni2005}. In contrast, during observations O4$^*$ and O4, the photon index increases to $\sim 2$ (Table \ref{Table: Model1S Parameters}), accompanied either by type-B QPOs (figure \ref{fig: PDS_seg2_O4*}) or by the absence of QPOs, consistent with the SIMS \citep{HomanBelloni2005}.

Further, we study the \textbf{rms} and \textbf{time lag} spectra of the observed QPOs in O4$^{*}$, and we observe that the rms increased with energy as reported by \citep{Mondal2023}. A similar increasing trend is observed for the time lags, which exhibits hard lags. \citet{Valentina2022} studied the rms and lag spectra of the type B QPO utilizing initial exposure of O4$^*$ and NICER. They used time-dependent Comptanization models \texttt{VKOMPTH} and \texttt{VKDUALDK} (see for \citealt{Bellavita2022} model details and references therein) to fit the simultaneous rms and lag spectra of the type B QPO and energy spectra of the GX 339-4. They require two physically-connected Comptonization regions to explain the radiative properties of the QPO. However, these models considers only the Compton scattering time-scale, which in general led to very large coronal sizes. Previous studies, such as \citet{Ma2023}, \citet{Rawat2023}, \citet{Rout2023}, \citet{Bellavita2025}, reported coronal size $>10^3$ km using this model, which is larger than expected. \citet{Valentina2022} also reported very large coronal size of $\sim18000$ km along with smaller corona with a size of $\sim 300$ km, and later they suggest the geometry of corona to be cylindrical rather than spherical which is the assumed geometry of this model. While \citet{Garg2020} assume the propagation delays between the disk and coronal variations and attempts to understand which radiative components, and in particular, the variation of which physical parameter, can model the observed rms and lag spectra of the QPO. The scheme has successfully described the milli-seconds time delays for low-frequency QPOs and broad features for several BHXBs (for e.g. \citealt{Garg2022, Nazma2023, Hitesh2024, Dhaka2024, Arbind2025}) while considering variations in the spectral parameters in most of the cases. Thereby, in this work, we intend to investigate whether such correlated variations between the disk and corona can also describe the variability features observed in the GX 339-4. 

\citet{Garg2020} demonstrated the application of this model to type C QPOs and their harmonics observed in GRS 1915+105. Their results suggest that the energy-dependent features of the QPOs can be explained by variations in the kT$_{\mathrm{in}}$, $R_{in}$, $\dot H$, and $\tau$, with short time delays between these parameters. \citet{Garg2022} apply the same model to type C QPOs observed in MAXI J1535–571. They find hard lags for QPO frequencies below 2 Hz and soft lags for frequencies above 2 Hz. Their results indicate that for QPOs with frequencies $<$ 2 Hz (hard lags), the variability originates in the disc and propagates inward towards the corona with millisecond time delays, whereas for the QPOs with frequencies $>$ 2 Hz (soft lags), the variations likely originate in the corona and then propagate outward towards the disc. \citet{Nazma2023} use this model to explore the radiative components responsible for the observed QPO and its harmonic in H 1743–322. Their findings are broadly consistent with earlier studies, showing that for the harmonic (associated with hard lags), the variations start in the disc and propagate inward towards the corona, while for the fundamental QPO (showing soft lags), the variability originates in the inner hot flow and propagates outward towards the disc. More recently, \citet{Arbind2025} applied this model to explain broad features, type C, and type B QPOs observed in MAXI J1803–298 across different spectral states, again reporting consistent results. Similarly, \citet{Dhaka2024} and \citet{Hitesh2024} used this model to interpret the broad features observed in the PDS of GRS 1915+105 and GX 339–4, respectively. \citet{Dhaka2024} report soft lags, while \citet{Hitesh2024} find hard lags, both supporting the same physical scenario inferred from the previous applications of this model.

We begin to fit the rms and time lag spectra of the observed type B QPOs during O4$^*$ by introducing the first-order variations in the inner disc temperature ($|\delta kT_{in}|$) and heating rate ($|\delta\dot{H}|$), with a phase-lag $\phi_{\dot{H}}$ between them (Model 1T), but we can not adequately fit the rms and lag spectra (Figure \ref{fig: LAG_RMS_Model_Fitting} \& \ref{fig: Chi2_comparision_M1_M2}). By introducing an additional variation in the disc normalization ($|\delta N_{disc}|$) and a phase-lag $\phi_{N_{disc}}$ with respect to $|\delta kT_{in}|$ (Model 2T) (Figure \ref{fig: LAG_RMS_Model_Fitting}), we reproduce the energy-dependent properties of the QPOs in most segments (Figure \ref{fig: Model_2_results_frms_lag}) with a $X^2_{red}$ close to 1 (Figure \ref{fig: Chi2_comparision_M1_M2}). From Figure \ref{fig: Model_2_results_frms_lag}, it is evident that the variations in $kT\textsubscript{in}$ are positive while those in $\dot H$ are negative with a phase lag zero. This indicates that when $kT\textsubscript{in}$ increases by $\delta kT\textsubscript{in}$, $\dot H$ decreases by $\delta \dot H$. Therefore, we observe that simultaneous variations in the $kT_{in}$, and $\dot H$ are the primary driver of the observed variability, while they are anti-correlated and after a $\sim10$ ms of time delay, these variations propagate to vary the $N_{dbb}$ parameter (Figure \ref{fig: Model_2_results_frms_lag}). Hence, we could also reproduce rms and lag spectra by introducing some variations in the disc and coronal parameters same as previous studies, however our results suggest that the variability originates in the the $kT_{in}$, and $\dot H$ simultaneously which then propagate to $N_{dbb}$ for the observed hard lags though previous studies reported variations get originated in the disc and propagate inwards towards the corona with some time delay. Although this model of \citet{Garg2020} represent the simplest picture of the complex phenomenon, it is remarkable that we as well as previous studies can fit the rms and lag spectra of the observed QPOs by just introducing first order variations in the spectral parameters. More studies in future like we present in this work, by applying this model to BHXBs, may give us better understanding about the role of radiative components in the energy-dependent features of the QPOs observed in BHXBs. Furthermore, in this analysis, we do not include reflection components for simplicity, as the applied model does not currently account for reflection effects. Incorporating reflection features into the model could provide a more complete understanding of the physical processes involved, highlighting the need for future model developments to include such complex spectral components.

In this work, we conducted a spectral and temporal study of the BHXB GX 339-4 during its 2021 outburst, and our key findings are as follows:

\begin{enumerate}
    \item We fit the combined energy spectrum of SXT and LAXPC in the energy range $0.7-25.0$ keV with \textbf{constant $\times$ tbabs $\times$ (gaussian + thcomp $\times$ diskbb)} (Model 1S) and assuming that the energy spectrum of BH LMXBs primarily consists of two components: a thermal component and a non-thermal component.

    \item The spectrum gets harder during O4, as we see the photon index and covering fraction both decreasing. Interestingly, the disc doesn’t change much since its temperature and apparent inner radius remain same, which is also confirmed by the stable unabsorbed flux in 0.1-3 keV energy range.

    \item Even though the soft X-ray flux stays steady, the hard X-ray flux keeps dropping, which suggests the corona is weakening over time. This likely explains the changes we see in the HID during O4$^*$ and O4.
   
    \item Study of the X-ray variability in the frequency domain through the PDS suggests that the QPO frequency increases from $\sim0.10$ Hz in O1 to $\sim0.18$ Hz in O2, and further to $\sim0.20$ Hz in O3, and decreases from $\sim 5.7$ Hz to $\sim 4.5$ Hz during O4$^*$.

    \item The rms of the QPO frequencies observed in the observation O4$^*$ increases with the energy and the lag spectra indicate the hard lags.
    
    \item To investigate the radiative component responsible for the observed QPO, we employ the generic scheme discussed in \citet{Garg2020} and \citet{Garg2022} to fit the rms and time lag spectra. We successfully reproduce the energy-dependent rms and time lag (Figure \ref{fig: LAG_RMS_Model_Fitting}). Interestingly, we find that the variations originate simultaneously in the inner disc temperature ($kT_{in}$) and the heating rate ($\dot H$); however, they are anti-correlated (Figure \ref{fig: Model_2_results_frms_lag}). After a time delay of $\sim 10$ ms, these variations propagate to the disc normalization ($N_{dbb}$).

\end{enumerate}

\section*{Acknowledgements}

We acknowledge the utilization of data obtained from the LAXPC and SXT instruments onboard the AstroSat satellite for this study. The analysis was done using the LAXPC software, SXT pipeline, and HEASoft tools. We thank the Indian Space Science Data Centre (ISSDC) for providing  access to the required data and resources. I, Vaibhav Sharma, also appreciate the help from the Inter-University Centre for Astronomy and Astrophysics (IUCAA) and am grateful for their warm hospitality during the visit, which significantly facilitated this work.

\section*{Data Availability}

The data used in this study is publicly available through the Indian Space Science Data Centre (ISSDC), which serves as the primary repository for the \textit{AstroSat} mission. Researchers can access the data via the ISSDC’s AstroBrowse platform at \url{http://astrobrowse.issdc.gov.in/astro_archive/archive}.



\bibliographystyle{mnras}
\bibliography{References} 

\begin{thebibliography}{}
\makeatletter
\relax
\def\mn@urlcharsother{\let\do\@makeother \do\$\do\&\do\#\do\^\do\_\do\%\do\~}
\def\mn@doi{\begingroup\mn@urlcharsother \@ifnextchar [ {\mn@doi@} {\mn@doi@[]}}
\def\mn@doi@[#1]#2{\def\@tempa{#1}\ifx\@tempa\@empty \href {http://dx.doi.org/#2} {doi:#2}\else \href {http://dx.doi.org/#2} {#1}\fi \endgroup}
\def\mn@eprint#1#2{\mn@eprint@#1:#2::\@nil}
\def\mn@eprint@arXiv#1{\href {http://arxiv.org/abs/#1} {{\tt arXiv:#1}}}
\def\mn@eprint@dblp#1{\href {http://dblp.uni-trier.de/rec/bibtex/#1.xml} {dblp:#1}}
\def\mn@eprint@#1:#2:#3:#4\@nil{\def\@tempa {#1}\def\@tempb {#2}\def\@tempc {#3}\ifx \@tempc \@empty \let \@tempc \@tempb \let \@tempb \@tempa \fi \ifx \@tempb \@empty \def\@tempb {arXiv}\fi \@ifundefined {mn@eprint@\@tempb}{\@tempb:\@tempc}{\expandafter \expandafter \csname mn@eprint@\@tempb\endcsname \expandafter{\@tempc}}}

\bibitem[\protect\citeauthoryear{Agrawal, Yadav, Antia  et~al.}{Agrawal et~al.}{2017}]{Agrawal2017}
Agrawal P.~C.,  Yadav J.~S.,  Antia H.~M.,   et~al., 2017, \mn@doi [Journal of Astrophysics and Astronomy] {10.1007/s12036-017-9451-z}, 38, 30

\bibitem[\protect\citeauthoryear{{Alabarta} et~al.,}{{Alabarta} et~al.}{2021}]{Alabarta2021}
{Alabarta} K.,  et~al., 2021, \mn@doi [\mnras] {10.1093/mnras/stab2241}, \href {https://ui.adsabs.harvard.edu/abs/2021MNRAS.507.5507A} {507, 5507}

\bibitem[\protect\citeauthoryear{Aneesha, Das, Katoch  \& Nandi}{Aneesha et~al.}{2024}]{Aneesha2024}
Aneesha U.,  Das S.,  Katoch T.~B.,   Nandi A.,  2024, \mn@doi [Monthly Notices of the Royal Astronomical Society] {10.1093/mnras/stae1753}, 532, 4486

\bibitem[\protect\citeauthoryear{Antia, Yadav, Agrawal  et~al.}{Antia et~al.}{2017}]{Antia2017}
Antia H.~M.,  Yadav J.~S.,  Agrawal P.~C.,   et~al., 2017, \mn@doi [The Astrophysical Journal Supplement Series] {10.3847/1538-4365/aa7a0e}, 231, 10

\bibitem[\protect\citeauthoryear{Antia et~al.}{Antia et~al.}{2021}]{Antia2021}
Antia H.~M.,  et~al., 2021, Journal of Astrophysics and Astronomy, 42, 32

\bibitem[\protect\citeauthoryear{{Arzoumanian} et~al.,}{{Arzoumanian} et~al.}{2014}]{Arzoumanian2014}
{Arzoumanian} Z.,  et~al., 2014, in {Takahashi} T.,  {den Herder} J.-W.~A.,   {Bautz} M.,  eds,  Society of Photo-Optical Instrumentation Engineers (SPIE) Conference Series Vol. 9144, Space Telescopes and Instrumentation 2014: Ultraviolet to Gamma Ray. p. 914420, \mn@doi{10.1117/12.2056811}

\bibitem[\protect\citeauthoryear{Bellavita, García, Méndez, Poutanen  \& Veledina}{Bellavita et~al.}{2022}]{Bellavita2022}
Bellavita C.,  García F.,  Méndez M.,  Poutanen J.,   Veledina A.,  2022, \mn@doi [Monthly Notices of the Royal Astronomical Society] {10.1093/mnras/stac1922}, 515, 2099

\bibitem[\protect\citeauthoryear{{Bellavita}, {M{\'e}ndez}, {Garc{\'\i}a}, {Ma}  \& {K{\"o}nig}}{{Bellavita} et~al.}{2025}]{Bellavita2025}
{Bellavita} C.,  {M{\'e}ndez} M.,  {Garc{\'\i}a} F.,  {Ma} R.,   {K{\"o}nig} O.,  2025, \mn@doi [\aap] {10.1051/0004-6361/202453092}, \href {https://ui.adsabs.harvard.edu/abs/2025A&A...696A.128B} {696, A128}

\bibitem[\protect\citeauthoryear{Belloni, Psaltis  \& van~der Klis}{Belloni et~al.}{2002}]{Belloni2002}
Belloni T.,  Psaltis D.,   van~der Klis M.,  2002, \mn@doi [The Astrophysical Journal] {10.1086/340290}, 572, 392

\bibitem[\protect\citeauthoryear{Belloni, Homan, Casella  \& et al.}{Belloni et~al.}{2005}]{Belloni2005}
Belloni T.,  Homan J.,  Casella P.,   et al. 2005, \mn@doi [Astronomy \& Astrophysics] {10.1051/0004-6361:20042457}, 440, 207

\bibitem[\protect\citeauthoryear{Belloni, Sanna  \& Méndez}{Belloni et~al.}{2012}]{Belloni2012}
Belloni T.~M.,  Sanna A.,   Méndez M.,  2012, \mn@doi [Monthly Notices of the Royal Astronomical Society] {10.1111/j.1365-2966.2012.21704.x}, 426, 1701

\bibitem[\protect\citeauthoryear{{Bhuvana}, {Radhika}  \& {Nandi}}{{Bhuvana} et~al.}{2021}]{2021ATel14455....1B}
{Bhuvana} G.~R.,  {Radhika} D.,   {Nandi} A.,  2021, The Astronomer's Telegram, \href {https://ui.adsabs.harvard.edu/abs/2021ATel14455....1B} {14455, 1}

\bibitem[\protect\citeauthoryear{{Casella}, {Belloni}  \& {Stella}}{{Casella} et~al.}{2005}]{Casella2005}
{Casella} P.,  {Belloni} T.,   {Stella} L.,  2005, \mn@doi [\apj] {10.1086/431174}, \href {https://ui.adsabs.harvard.edu/abs/2005ApJ...629..403C} {629, 403}

\bibitem[\protect\citeauthoryear{Chakrabarti \& Titarchuk}{Chakrabarti \& Titarchuk}{1995}]{Chakrabarti1995}
Chakrabarti S.,  Titarchuk L.~G.,  1995, \mn@doi [The Astrophysical Journal] {10.1086/176610}, 455, 623

\bibitem[\protect\citeauthoryear{Chakrabarti, Debnath, Nandi  \& Pal}{Chakrabarti et~al.}{2008}]{Chakrabarti2008}
Chakrabarti S.~K.,  Debnath D.,  Nandi A.,   Pal P.~S.,  2008, \mn@doi [Astronomy \& Astrophysics] {10.1051/0004-6361:200810136}, 489, L41

\bibitem[\protect\citeauthoryear{Chand, Dewangan, Zdziarski, Bhattacharya, Mithun  \& Vadawale}{Chand et~al.}{2024}]{Chand2024}
Chand S.,  Dewangan G.~C.,  Zdziarski A.~A.,  Bhattacharya D.,  Mithun N. P.~S.,   Vadawale S.~V.,  2024, \mn@doi [The Astrophysical Journal] {10.3847/1538-4357/ad5a88}, 972, 20

\bibitem[\protect\citeauthoryear{Deegan, Combet  \& Wynn}{Deegan et~al.}{2009}]{Deegan2009}
Deegan P.,  Combet C.,   Wynn G.~A.,  2009, \mn@doi [Monthly Notices of the Royal Astronomical Society] {10.1111/j.1365-2966.2009.15573.x}, 400, 1337

\bibitem[\protect\citeauthoryear{Dhaka, Misra, Jain  \& Yadav}{Dhaka et~al.}{2024}]{Dhaka2024}
Dhaka R.,  Misra R.,  Jain P.,   Yadav J.~S.,  2024, \mn@doi [The Astrophysical Journal] {10.3847/1538-4357/ad67e4}, 974, 90

\bibitem[\protect\citeauthoryear{{Done}, {Gierli{\'n}ski}  \& {Kubota}}{{Done} et~al.}{2007}]{Done2007}
{Done} C.,  {Gierli{\'n}ski} M.,   {Kubota} A.,  2007, \mn@doi [\aapr] {10.1007/s00159-007-0006-1}, \href {https://ui.adsabs.harvard.edu/abs/2007A&ARv..15....1D} {15, 1}

\bibitem[\protect\citeauthoryear{Dubus, Hameury  \& Lasota}{Dubus et~al.}{2001}]{Dubus2001}
Dubus G.,  Hameury J.~M.,   Lasota J.~P.,  2001, \mn@doi [Astronomy \& Astrophysics] {10.1051/0004-6361:20010632}, 373, 251

\bibitem[\protect\citeauthoryear{Fabian, Rees, Stella  \& White}{Fabian et~al.}{1989}]{Fabian1989}
Fabian A.~C.,  Rees M.~J.,  Stella L.,   White N.~E.,  1989, Monthly Notices of the Royal Astronomical Society, 238, 729

\bibitem[\protect\citeauthoryear{Garcia, Tomsick, Harrison, Connors  \& Mastroserio}{Garcia et~al.}{2021}]{Garcia2021}
Garcia J.,  Tomsick J.,  Harrison F.,  Connors R.,   Mastroserio G.,  2021, The Astronomer’s Telegram, 14352, 1

\bibitem[\protect\citeauthoryear{Garg, Misra  \& Sen}{Garg et~al.}{2020}]{Garg2020}
Garg A.,  Misra R.,   Sen S.,  2020, Monthly Notices of the Royal Astronomical Society, 498, 2757

\bibitem[\protect\citeauthoryear{Garg, Misra  \& Sen}{Garg et~al.}{2022}]{Garg2022}
Garg A.,  Misra R.,   Sen S.,  2022, Monthly Notices of the Royal Astronomical Society, 514, 3285

\bibitem[\protect\citeauthoryear{{Gendreau} et~al.,}{{Gendreau} et~al.}{2016}]{Gendreau2016}
{Gendreau} K.~C.,  et~al., 2016, in {den Herder} J.-W.~A.,  {Takahashi} T.,   {Bautz} M.,  eds,  Society of Photo-Optical Instrumentation Engineers (SPIE) Conference Series Vol. 9905, Space Telescopes and Instrumentation 2016: Ultraviolet to Gamma Ray. p. 99051H, \mn@doi{10.1117/12.2231304}

\bibitem[\protect\citeauthoryear{Haardt \& Maraschi}{Haardt \& Maraschi}{1993}]{Haardt1993}
Haardt F.,  Maraschi L.,  1993, \mn@doi [The Astrophysical Journal] {10.1086/173020}, 413, 507

\bibitem[\protect\citeauthoryear{Harrison, Craig, Christensen, Hailey, Zhang, Boggs, Stern  et~al.}{Harrison et~al.}{2013}]{Harrison2013}
Harrison F.~A.,  Craig W.~W.,  Christensen F.~E.,  Hailey C.~J.,  Zhang W.~W.,  Boggs S.~E.,  Stern D.,   et~al., 2013, \mn@doi [ApJ] {10.1088/0004-637X/770/2/103}, 770, 103

\bibitem[\protect\citeauthoryear{Heida, Jonker, Torres  \& Chiavassa}{Heida et~al.}{2017}]{Heida2017}
Heida M.,  Jonker P.~G.,  Torres M. A.~P.,   Chiavassa A.,  2017, \mn@doi [The Astrophysical Journal] {10.3847/1538-4357/aa85df}, 846, 132

\bibitem[\protect\citeauthoryear{{Homan} \& {Belloni}}{{Homan} \& {Belloni}}{2005}]{HomanBelloni2005}
{Homan} J.,  {Belloni} T.,  2005, \mn@doi [\apss] {10.1007/s10509-005-1197-4}, \href {https://ui.adsabs.harvard.edu/abs/2005Ap&SS.300..107H} {300, 107}

\bibitem[\protect\citeauthoryear{{Homan}, {Wijnands}, {van der Klis}, {Belloni}, {van Paradijs}, {Klein-Wolt}, {Fender}  \& {M{\'e}ndez}}{{Homan} et~al.}{2001}]{Homan2001}
{Homan} J.,  {Wijnands} R.,  {van der Klis} M.,  {Belloni} T.,  {van Paradijs} J.,  {Klein-Wolt} M.,  {Fender} R.,   {M{\'e}ndez} M.,  2001, \mn@doi [\apjs] {10.1086/318954}, \href {https://ui.adsabs.harvard.edu/abs/2001ApJS..132..377H} {132, 377}

\bibitem[\protect\citeauthoryear{Husain, Mudambi, Garg, Misra, Sen  \& Maqbool}{Husain et~al.}{2021}]{Husain2021}
Husain N.,  Mudambi S.~P.,  Garg A.,  Misra R.,  Sen S.,   Maqbool B.,  2021, The Astronomer’s Telegram, 14400, 1

\bibitem[\protect\citeauthoryear{Hussain, Garg, Misra  \& Sen}{Hussain et~al.}{2023}]{Nazma2023}
Hussain N.,  Garg A.,  Misra R.,   Sen S.,  2023, Monthly Notices of the Royal Astronomical Society, 525, 4515

\bibitem[\protect\citeauthoryear{{Hynes}, {Steeghs}, {Casares}, {Charles}  \& {O'Brien}}{{Hynes} et~al.}{2003}]{Hynes2003}
{Hynes} R.~I.,  {Steeghs} D.,  {Casares} J.,  {Charles} P.~A.,   {O'Brien} K.,  2003, \mn@doi [\apjl] {10.1086/368108}, \href {https://ui.adsabs.harvard.edu/abs/2003ApJ...583L..95H} {583, L95}

\bibitem[\protect\citeauthoryear{Ingram \& Motta}{Ingram \& Motta}{2019}]{Ingram2019}
Ingram A.~R.,  Motta S.~E.,  2019, \mn@doi [New Astronomy Reviews] {10.1016/j.newar.2020.101524}, 85, 101524

\bibitem[\protect\citeauthoryear{Ingram, Done  \& Fragile}{Ingram et~al.}{2009}]{Ingram2009}
Ingram A.,  Done C.,   Fragile P.~C.,  2009, \mn@doi [Monthly Notices of the Royal Astronomical Society] {10.1111/j.1745-3933.2009.00697.x}, 397, L101

\bibitem[\protect\citeauthoryear{Jana, Chatterjee, Chang, Naik  \& Mondal}{Jana et~al.}{2024}]{Jana2024}
Jana A.,  Chatterjee D.,  Chang H.-K.,  Naik S.,   Mondal S.,  2024, \mn@doi [Monthly Notices of the Royal Astronomical Society] {10.1093/mnras/stad3192}, 527, 2128

\bibitem[\protect\citeauthoryear{{Kaastra} \& {Bleeker}}{{Kaastra} \& {Bleeker}}{2016}]{Kaastra...2016}
{Kaastra} J.~S.,  {Bleeker} J. A.~M.,  2016, \mn@doi [\aap] {10.1051/0004-6361/201527395}, \href {https://ui.adsabs.harvard.edu/abs/2016A&A...587A.151K} {587, A151}

\bibitem[\protect\citeauthoryear{Karpouzas, Méndez, García, Poutanen  \& Veledina}{Karpouzas et~al.}{2020}]{Karpouzas2020}
Karpouzas K.,  Méndez M.,  García F.,  Poutanen J.,   Veledina A.,  2020, \mn@doi [Monthly Notices of the Royal Astronomical Society] {10.1093/mnras/stz3502}, 492, 1399

\bibitem[\protect\citeauthoryear{Liu, Ji, Bambi, Jain, Misra, Rawat, Yadav  \& Zhang}{Liu et~al.}{2021}]{Liu2021}
Liu H.,  Ji L.,  Bambi C.,  Jain P.,  Misra R.,  Rawat D.,  Yadav J.,   Zhang Y.,  2021, \mn@doi [The Astrophysical Journal] {10.3847/1538-4357/abdb3d}, 909, 63

\bibitem[\protect\citeauthoryear{{Ma}, {M{\'e}ndez}, {Garc{\'\i}a}, {Sai}, {Zhang}  \& {Zhang}}{{Ma} et~al.}{2023}]{Ma2023}
{Ma} R.,  {M{\'e}ndez} M.,  {Garc{\'\i}a} F.,  {Sai} N.,  {Zhang} L.,   {Zhang} Y.,  2023, \mn@doi [\mnras] {10.1093/mnras/stad2284}, \href {https://ui.adsabs.harvard.edu/abs/2023MNRAS.525..854M} {525, 854}

\bibitem[\protect\citeauthoryear{Maqbool, Mudambi, Misra, Yadav  \& Jain}{Maqbool et~al.}{2019}]{Maqbool2019}
Maqbool B.,  Mudambi S.~P.,  Misra R.,  Yadav J.~S.,   Jain P.,  2019, \mn@doi [Monthly Notices of the Royal Astronomical Society] {10.1093/mnras/stz930}, 486, 2964

\bibitem[\protect\citeauthoryear{Markert, Canizares, Clark  \& et al.}{Markert et~al.}{1973}]{Markert1973}
Markert T.~H.,  Canizares C.~R.,  Clark G.~W.,   et al. 1973, \mn@doi [The Astrophysical Journal Letters] {10.1086/181290}, 184, L67

\bibitem[\protect\citeauthoryear{Matt, Perola  \& Piro}{Matt et~al.}{1991}]{Matt1991}
Matt G.,  Perola G.~C.,   Piro L.,  1991, Astronomy \& Astrophysics, 247, 25

\bibitem[\protect\citeauthoryear{Misra \& Mandal}{Misra \& Mandal}{2013}]{Misra2013}
Misra R.,  Mandal S.,  2013, \mn@doi [The Astrophysical Journal] {10.1088/0004-637X/779/1/71}, 779, 71

\bibitem[\protect\citeauthoryear{Mitsuda, Inoue, Koyama  et~al.}{Mitsuda et~al.}{1984}]{mitsuda1984}
Mitsuda K.,  Inoue H.,  Koyama K.,   et~al., 1984, Publications of the Astronomical Society of Japan, 36, 741

\bibitem[\protect\citeauthoryear{Mondal, Salgundi, Chatterjee, Jana, Chang  \& Naik}{Mondal et~al.}{2023}]{Mondal2023}
Mondal S.,  Salgundi A.,  Chatterjee D.,  Jana A.,  Chang H.-K.,   Naik S.,  2023, \mn@doi [Monthly Notices of the Royal Astronomical Society] {10.1093/mnras/stad3079}, 526, 4718

\bibitem[\protect\citeauthoryear{{Motta}, {Casella}, {Henze}, {Mu{\ n}oz-Darias}, {Sanna}, {Fender}  \& {Belloni}}{{Motta} et~al.}{2015}]{Motta2015}
{Motta} S.~E.,  {Casella} P.,  {Henze} M.,  {Mu{\ n}oz-Darias} T.,  {Sanna} A.,  {Fender} R.,   {Belloni} T.,  2015, \mn@doi [\mnras] {10.1093/mnras/stu2579}, \href {https://ui.adsabs.harvard.edu/abs/2015MNRAS.447.2059M} {447, 2059}

\bibitem[\protect\citeauthoryear{Motta, Belloni, Stella, Pappas, Casares, Muñoz-Darias, Torres  \& Yanes-Rizo}{Motta et~al.}{2022}]{Motta2022}
Motta S.~E.,  Belloni T.,  Stella L.,  Pappas G.,  Casares J.,  Muñoz-Darias T.,  Torres M. A.~P.,   Yanes-Rizo I.~V.,  2022, \mn@doi [Monthly Notices of the Royal Astronomical Society] {10.1093/mnras/stac2770}, 517, 1469

\bibitem[\protect\citeauthoryear{Méndez, Altamirano, Belloni  \& Sanna}{Méndez et~al.}{2013}]{Mendez2013}
Méndez M.,  Altamirano D.,  Belloni T.,   Sanna A.,  2013, \mn@doi [Monthly Notices of the Royal Astronomical Society] {10.1093/mnras/stt1431}, 435, 2132

\bibitem[\protect\citeauthoryear{Nowak}{Nowak}{2000}]{Nowak2000}
Nowak M.~A.,  2000, \mn@doi [Monthly Notices of the Royal Astronomical Society] {10.1046/j.1365-8711.2000.03779.x}, 318, 361

\bibitem[\protect\citeauthoryear{Parker, Tomsick, Kennea  \& et al.}{Parker et~al.}{2016}]{Parker2016}
Parker M.~L.,  Tomsick J.~A.,  Kennea J.~A.,   et al. 2016, \mn@doi [The Astrophysical Journal Letters] {10.3847/2041-8205/821/1/L6}, 821, L6

\bibitem[\protect\citeauthoryear{Peirano, Méndez, García  \& Belloni}{Peirano et~al.}{2022}]{Valentina2022}
Peirano V.,  Méndez M.,  García F.,   Belloni T.,  2022, Monthly Notices of the Royal Astronomical Society, 519, 1336

\bibitem[\protect\citeauthoryear{{Pradhan}, {Garg}, {Misra}  \& {Sarkar}}{{Pradhan} et~al.}{2025}]{Arbind2025}
{Pradhan} A.,  {Garg} A.,  {Misra} R.,   {Sarkar} B.,  2025, \mn@doi [\apj] {10.3847/1538-4357/adfa72}, \href {https://ui.adsabs.harvard.edu/abs/2025ApJ...991...93P} {991, 93}

\bibitem[\protect\citeauthoryear{{Rawat} et~al.,}{{Rawat} et~al.}{2023}]{Rawat2023}
{Rawat} D.,  et~al., 2023, \mn@doi [\mnras] {10.1093/mnras/stad126}, \href {https://ui.adsabs.harvard.edu/abs/2023MNRAS.520..113R} {520, 113}

\bibitem[\protect\citeauthoryear{Remillard \& McClintock}{Remillard \& McClintock}{2006}]{Remillard2006}
Remillard R.~A.,  McClintock J.~E.,  2006, \mn@doi [Annual Review of Astronomy and Astrophysics] {10.48550/arXiv.astro-ph/0606352}, 44, 49

\bibitem[\protect\citeauthoryear{{Rout}, {M{\'e}ndez}  \& {Garc{\'\i}a}}{{Rout} et~al.}{2023}]{Rout2023}
{Rout} S.~K.,  {M{\'e}ndez} M.,   {Garc{\'\i}a} F.,  2023, \mn@doi [\mnras] {10.1093/mnras/stad2321}, \href {https://ui.adsabs.harvard.edu/abs/2023MNRAS.525..221R} {525, 221}

\bibitem[\protect\citeauthoryear{Shakura \& Sunyaev}{Shakura \& Sunyaev}{1973}]{Shakura1973}
Shakura N.~I.,  Sunyaev R.~A.,  1973, Astronomy \& Astrophysics, 24, 337

\bibitem[\protect\citeauthoryear{{Singh} et~al.,}{{Singh} et~al.}{2014}]{Singh2014}
{Singh} K.~P.,  et~al., 2014, in {Takahashi} T.,  {den Herder} J.-W.~A.,   {Bautz} M.,  eds,  Society of Photo-Optical Instrumentation Engineers (SPIE) Conference Series Vol. 9144, Space Telescopes and Instrumentation 2014: Ultraviolet to Gamma Ray. p. 91441S, \mn@doi{10.1117/12.2062667}

\bibitem[\protect\citeauthoryear{Singh, Stewart, Chandra  et~al.}{Singh et~al.}{2016}]{Singh2016}
Singh K.~P.,  Stewart G.~C.,  Chandra S.,   et~al., 2016, in den Herder J.-W.~A.,  Takahashi T.,   Bautz M.,  eds,  Society of Photo-Optical Instrumentation Engineers (SPIE) Conference Series Vol. 9905, Space Telescopes and Instrumentation 2016: Ultraviolet to Gamma Ray. p. 99051E, \mn@doi{10.1117/12.2235309}

\bibitem[\protect\citeauthoryear{Singh, Stewart, Westergaard  et~al.}{Singh et~al.}{2017}]{Singh2017}
Singh K.~P.,  Stewart G.~C.,  Westergaard N.~J.,   et~al., 2017, \mn@doi [Journal of Astrophysics and Astronomy] {10.1007/s12036-017-9448-7}, 38, 29

\bibitem[\protect\citeauthoryear{{Sobczak}, {Remillard}, {Muno}  \& {McClintock}}{{Sobczak} et~al.}{2000}]{Sobczak2000}
{Sobczak} G.~J.,  {Remillard} R.~A.,  {Muno} M.~P.,   {McClintock} J.~E.,  2000, \mn@doi [arXiv e-prints] {10.48550/arXiv.astro-ph/0004215}, \href {https://ui.adsabs.harvard.edu/abs/2000astro.ph..4215S} {pp astro--ph/0004215}

\bibitem[\protect\citeauthoryear{Sreehari, Iyer, Radhika, Nandi  \& Mandal}{Sreehari et~al.}{2019}]{Sreehari2019}
Sreehari H.,  Iyer N.,  Radhika D.,  Nandi A.,   Mandal S.,  2019, \mn@doi [Advances in Space Research] {10.1016/j.asr.2018.10.042}, 63, 1374

\bibitem[\protect\citeauthoryear{Stella \& Vietri}{Stella \& Vietri}{1997}]{Stella1997}
Stella L.,  Vietri M.,  1997, \mn@doi [The Astrophysical Journal] {10.1086/311075}, 492, L59

\bibitem[\protect\citeauthoryear{Stella, Vietri  \& Morsink}{Stella et~al.}{1999}]{Stella1999}
Stella L.,  Vietri M.,   Morsink S.~M.,  1999, \mn@doi [The Astrophysical Journal] {10.1086/312291}, 524, L63

\bibitem[\protect\citeauthoryear{Tanenia, Garg, Misra  \& Sen}{Tanenia et~al.}{2024}]{Hitesh2024}
Tanenia H.,  Garg A.,  Misra R.,   Sen S.,  2024, The Astrophysical Journal, 975, 190

\bibitem[\protect\citeauthoryear{{Tetarenko}, {Sivakoff}, {Heinke}  \& {Gladstone}}{{Tetarenko} et~al.}{2016}]{Tetarenko2016}
{Tetarenko} B.~E.,  {Sivakoff} G.~R.,  {Heinke} C.~O.,   {Gladstone} J.~C.,  2016, \mn@doi [\apjs] {10.3847/0067-0049/222/2/15}, \href {https://ui.adsabs.harvard.edu/abs/2016ApJS..222...15T} {222, 15}

\bibitem[\protect\citeauthoryear{Titarchuk \& Osherovich}{Titarchuk \& Osherovich}{2000}]{Titarchuk2000}
Titarchuk L.,  Osherovich V.,  2000, \mn@doi [The Astrophysical Journal Letters] {10.1086/312682}, 542, L111

\bibitem[\protect\citeauthoryear{Wang et~al.}{Wang et~al.}{2021}]{Wang2021}
Wang J.,  et~al., 2021, The Astronomer’s Telegram, 14384, 1

\bibitem[\protect\citeauthoryear{{Wijnands}, {Homan}  \& {van der Klis}}{{Wijnands} et~al.}{1999}]{Wijnands1999}
{Wijnands} R.,  {Homan} J.,   {van der Klis} M.,  1999, \mn@doi [\apjl] {10.1086/312365}, \href {https://ui.adsabs.harvard.edu/abs/1999ApJ...526L..33W} {526, L33}

\bibitem[\protect\citeauthoryear{Wilms, Allen  \& McCray}{Wilms et~al.}{2000}]{Wilms2000}
Wilms J.,  Allen A.,   McCray R.,  2000, \mn@doi [The Astrophysical Journal] {10.1086/317016}, 542, 914

\bibitem[\protect\citeauthoryear{Yadav, Misra, Verdhan~Chauhan  et~al.}{Yadav et~al.}{2016a}]{Yadav2016b}
Yadav J.~S.,  Misra R.,  Verdhan~Chauhan J.,   et~al., 2016a, \mn@doi [The Astrophysical Journal] {10.3847/0004-637X/833/1/27}, 833, 27

\bibitem[\protect\citeauthoryear{Yadav, Agrawal, Antia  et~al.}{Yadav et~al.}{2016b}]{Yadav2016a}
Yadav J.~S.,  Agrawal P.~C.,  Antia H.~M.,   et~al., 2016b, in den Herder J.-W.~A.,  Takahashi T.,   Bautz M.,  eds,  Society of Photo-Optical Instrumentation Engineers (SPIE) Conference Series Vol. 9905, Space Telescopes and Instrumentation 2016: Ultraviolet to Gamma Ray. p. 99051D, \mn@doi{10.1117/12.2231857}

\bibitem[\protect\citeauthoryear{Zdziarski, Ziółkowski  \& Mikołajewska}{Zdziarski et~al.}{2019}]{Zdziarski2019}
Zdziarski A.~A.,  Ziółkowski J.,   Mikołajewska J.,  2019, \mn@doi [Monthly Notices of the Royal Astronomical Society] {10.1093/mnras/stz1787}, 488, 1026

\bibitem[\protect\citeauthoryear{Zdziarski, Szanecki, Poutanen, Gierliński  \& Biernacki}{Zdziarski et~al.}{2020}]{Zdziarski2020}
Zdziarski A.~A.,  Szanecki M.,  Poutanen J.,  Gierliński M.,   Biernacki P.,  2020, \mn@doi [Monthly Notices of the Royal Astronomical Society] {10.1093/mnras/staa159}, 492, 5234

\bibitem[\protect\citeauthoryear{{Zhang} et~al.,}{{Zhang} et~al.}{2020}]{Zhang2020}
{Zhang} S.-N.,  et~al., 2020, \mn@doi [Science China Physics, Mechanics, and Astronomy] {10.1007/s11433-019-1432-6}, \href {https://ui.adsabs.harvard.edu/abs/2020SCPMA..6349502Z} {63, 249502}

\makeatother
\end{thebibliography}





\bsp	
\label{lastpage}
\end{document}